\documentclass[epj,final]{svjour}

\usepackage{epsfig}
\usepackage{booktabs}
\usepackage{amsfonts}

\newcommand{\Lambdareaktion}{$\gamma\,p\,\rightarrow\,K^+\,\Lambda~$}
\newcommand{\Lambdareaktionvorkomma}{$\gamma\,p\,\rightarrow\,K^+\,\Lambda$}
\newcommand{\Sigmanreaktion}{$\gamma\,p\,\rightarrow\,K^+\,\Sigma^0~$}
\newcommand{\Sigmanreaktionvorkomma}{$\gamma\,p\,\rightarrow\,K^+\,\Sigma^0$}
\newcommand{\Zweipireaktion}{$\gamma\,p\,\rightarrow\,p\,\pi^+\,\pi^-$}
\newcommand{\Dreipireaktion}{$\gamma\,p\,\rightarrow\,p\,\pi^+\,\pi^-\,\pi^0$}
\newcommand{\Lambdapinull}{$\gamma\,p\,\rightarrow\,K^+\,\Lambda\,\pi^0~$}
\newcommand{\Lambdapinullvorkomma}{$\gamma\,p\,\rightarrow\,K^+\,\Lambda\,\pi^0$}
\newcommand{\Lambdaklspip}{$\gamma\,p\,\rightarrow\,K^0\,\Lambda\,\pi^+~$}

\newcommand{\coskaoncmsohne}{$cos(\theta^{\,cms}_{K^+})$~}
\newcommand{\coskaoncmsvorkomma}{$cos(\theta^{\,cms}_{K^+})$}
\newcommand{\Lambdazerfall}{$\Lambda\,\rightarrow\,p\,\pi^-$~}
\newcommand{\Lambdazerfallvorkomma}{$\Lambda\,\rightarrow\,p\,\pi^-$}

\newcommand{\Neutrondreipivorkomma}{$\gamma\,p\,\rightarrow\,n\,\pi^+\,\pi^-\,\pi^+$}
\newcommand{\Sigmazerfall}{$\Sigma^0\,\rightarrow\,\Lambda\,\gamma$~}
\newcommand{\Sigmazerfallvorkomma}{$\Sigma^0\,\rightarrow\,\Lambda\,\gamma$}
\def\miskaon{$m_{\gamma\,p\,-\,K^+}$~}
\def\miskaonvorkomma{$m_{\gamma\,p\,-\,K^+}$}
\def\mvxfit{$m_\Lambda$~}
\def\mvxfitvorkomma{$m_\Lambda$}

\begin{document}

\title{Measurement of \boldmath \Lambdareaktionvorkomma\ \unboldmath and \boldmath \Sigmanreaktionvorkomma\ \unboldmath
at photon energies up to 2.6~GeV\thanks{This work is supported in part by the Deutsche
Forschungsgemeinschaft (DFG) (SPP 1034 KL 980/2-3)}}
\titlerunning{Measurement of \Lambdareaktionvorkomma\ and \Sigmanreaktionvorkomma\ at photon energies up to 2.6~GeV}

\author{K.-H.\ Glander\inst{1}\fnmsep\thanks{Part of doctoral thesis \cite{glander}}\and
J.\ Barth\inst{1}\and W.\ Braun\inst{1,3}\and J.\ Hannappel\inst{1,3}\and N.\,J{\"o}pen\inst{1}\and
F.\ Klein\inst{1}\and E.\ Klempt\inst{2}\and R.\ Lawall\inst{1}\and J.\ Link\inst{2,3}\and
D.\ Menze\inst{1}\and W.\,Neuerburg\inst{1,3}\and M.\ Ostrick\inst{1}\and E.\ Paul\inst{1}\and
I.\ Schulday\inst{1}\and W.\,J.\,Schwille\inst{1}\and H.\,v.\ Pee\inst{2,3}\and
F.\,W.\ Wieland\inst{1}\and J.\ Wi{\ss}kirchen\inst{1,3}\and C.\ Wu\inst{1}}
\authorrunning{K.-H.\ Glander {\it et al.}}

\institute{
     Physikalisches Institut, Bonn University, Bonn, Germany
\and Helmholtz-Institut f{\"u}r Strahlen- und Kernphysik, Bonn University, Bonn, Germany
\and No longer working at this experiment}

\mail{klein@physik.uni-bonn.de}

\abstract
{The reactions \Lambdareaktionvorkomma\ and \Sigmanreaktionvorkomma\ were measured in the energy
range from threshold up to a photon energy of 2.6~GeV. The data
were taken with the SAPHIR detector at the electron stretcher facility, ELSA.
Results on cross sections and hyperon polarizations are presented
as a function of kaon production angle and photon energy.
The total cross section for $\Lambda$ production rises steeply with energy
close to threshold, whereas the $\Sigma^0$ cross section rises slowly to a maximum at about
$E_\gamma\,=\,1.45~\mbox{GeV}$.
Cross sections together with their angular decompositions
into Legendre polynomials suggest
contributions from resonance production for both reactions.
%The $K^+\,\Lambda$ differential cross section is enhanced for
%backward produced kaons at $E_\gamma \approx 1.45~\mbox{GeV}$.
%This might be interpreted as contribution of a {\it missing resonance}
%$D_{13}(1895)$ which has been proposed by model calculations.
In general, the induced polarization of $\Lambda$ has negative values in the kaon forward direction
and positive values in the backward direction. The magnitude varies with energy.
The polarization of $\Sigma^0$ follows a similar angular and energy dependence
as that of $\Lambda$, but with opposite sign.
}

\maketitle

\section{Introduction}

The production of open strangeness in photon-induced hadronic
processes on protons provides a tool to investigate the dynamical role
of flavours, since a strange quark-antiquark pair
is created. This study contributes
to the search for missing non-strange resonances which decay into strange particle
pairs (see e.g.~\cite{capstick,loering}).\\
We report on measurements of total and differential
\linebreak[4]
cross sections and hyperon
polarizations in the reaction
channels~\Lambdareaktionvorkomma\ (\Lambdazerfallvorkomma)~and~\Sigmanreaktionvorkomma
\linebreak[4]
(\Sigmazerfallvorkomma; \Lambdazerfallvorkomma)
between threshold and a photon energy of 2.6~GeV.\\
The data were taken with the magnetic multiparticle detector, SAPHIR \cite{schwille},
at the 3.5~GeV electron stretcher accelerator, ELSA \cite{husmann},
with electron beam energies of 2.8~GeV and 2.6~GeV.\\
Results from SAPHIR on the reactions \Lambdareaktionvorkomma\ and \Sigmanreaktionvorkomma, based on
30 million triggers taken in 1992-94 with photon energies up to 2.0~GeV, were already
published \cite{tran}.\\
The new results reported here come from four data taking periods in
1997/98 where 180 million triggers were collected with tagged photons
in an extended energy range up to 2.65~GeV.\\
The data are available via the internet\footnote{http://saphir.physik.uni-bonn.de/saphir/publications.html}.

\section{Experiment and event reconstruction}
\label{sec:datataking}

\begin{figure}
\vspace{-0.15cm}
\includegraphics[clip,angle=270,width=0.47\textwidth]{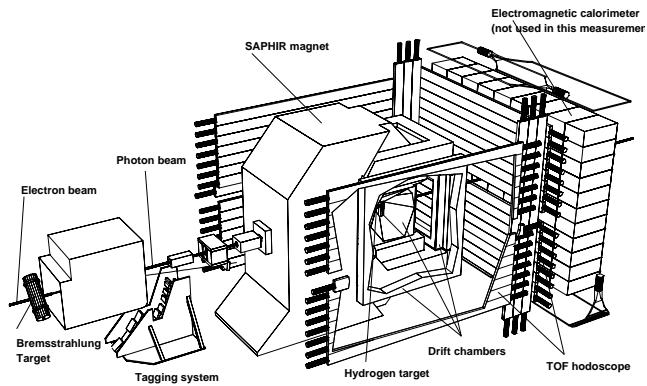}
\caption{Sketch of the SAPHIR detector.}
\label{pic:saphir}
\vspace{-0.3cm}
\end{figure}
The magnetic multiparticle detector, SAPHIR \cite{schwille}, is
shown schematically in Fig.~\ref{pic:saphir}.
The extracted electron beam was directed on a radiator target to
provide a photon beam which was energy-tagged over the range from about
\linebreak[4]
0.85~GeV to 2.65~GeV (0.8~GeV to 2.45~GeV for electrons of 2.6~GeV).
The data taking was based on a trigger
defined by a coincidence of signals from the scattered electron in the
tagging system with at least two charged particles in the scintillator
hodoscopes of SAPHIR and no signal from a beam veto counter of non-interacting photons.\\
The SAPHIR detector was upgraded for this data taking.
A new tagging system with a larger photon energy range \cite{topas2,barth2}
was installed.
The tagging system and the photon beam veto counter were used to measure
the photon flux.
The tagging system comprises 14 scintillation counters
for triggering and time definition and 2 multi-wire proportional
chambers defining 703 energy channels. Each scintillation counter is
connected to a scaler. Approximately every 0.4 seconds a minimum bias
trigger defined by an electron that hits a counter in the tagging
system starts a full read-out of the SAPHIR data including the
scalers. This event sample was used to calculate a normalisation
factor, based on $N_e$ as the number of hits in all
tagging scintillators and $N_{\gamma_i}$ as the number of
coincidences between the photon veto counter, the energy channel $i$,
and the associated tagging scintillator.  Multiplying the ratio
$N_{\gamma_i}$/$N_e$ with the total number of
hits in the tagging scintillators for a run period provides the photon
flux for the channel $i$.  Thus, since data taking and
flux normalisation take place simultaneously, any inefficiency of the
tagger, even if it varies during the data acquisition, is
automatically taken into account.\\
A planar drift chamber with the same
hexagonal drift cell structure as the central drift chamber has been
added downstream of the central drift chamber in order to improve the track
reconstruction in forward direction \cite{glander}.
The drift chambers were filled with a gas mixture of
neon (63.75\%), helium (21.25\%) and isobutane (15\%).
By including the hits of the forward
drift chamber in the track fit the momentum resolution for tracks
crossing the forward drift chamber was improved on average by a factor of five \cite{glander}.
The momentum resolution $\Delta p/p$ for a 300\,MeV/c particle measured in the central drift
chamber is about 2.5\,\%. The forward drift chamber
ameliorates the momentum resolution to about $1$\,\%. For 1\,GeV/c particles
$\Delta p/p$ is 7.5\,\% in the central drift chamber and 1.5\,\%
with assistance of the forward drift chamber \cite{glander}.\\
Events were reconstructed from measurements of the incident photon in the tagging system
and charged particles in the drift chamber system.

\section{Data selection}
\label{sec:selection}

\begin{figure}[htb]
\vspace{-1.2cm}
\centerline{
\includegraphics[clip,width=0.53\textwidth]{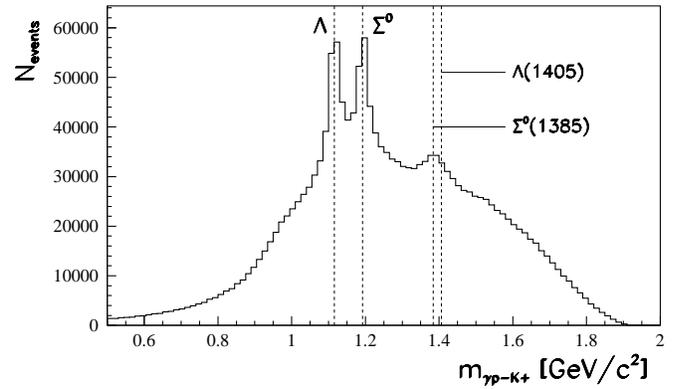}}
\vspace{-0.1cm}
\caption{Distribution of the missing mass \miskaonvorkomma\ after 
the selection based on the event topology.}
\vspace{-0.1cm}
\label{pic:minv0}
\end{figure}

\begin{figure}[htb]
\vspace{-1.3cm}
\centerline{
\includegraphics[clip,width=0.53\textwidth]{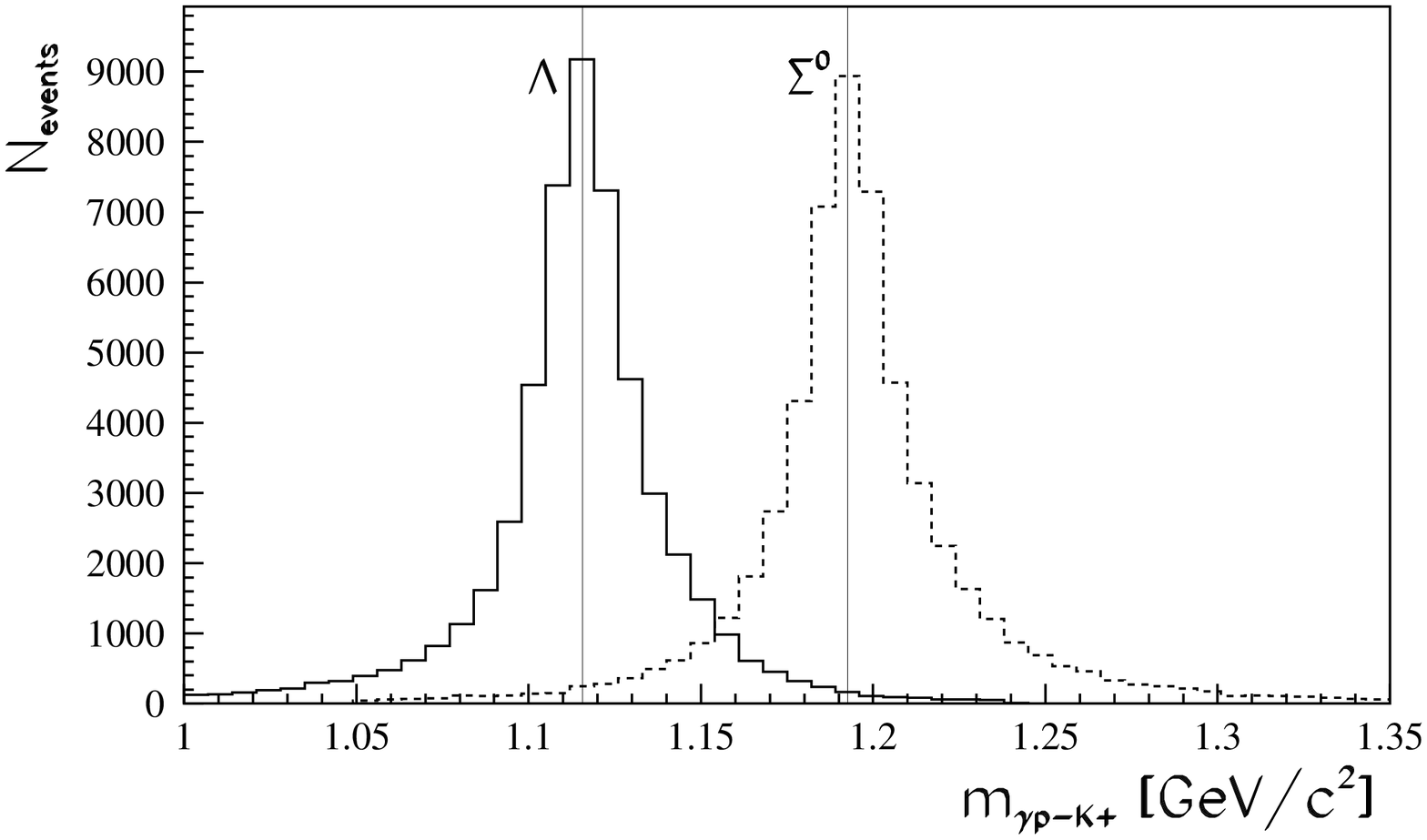}}
\vspace{-0.1cm}
\caption{Distribution of the missing mass \miskaon for events from the reactions \Lambdareaktion
and \Sigmanreaktionvorkomma\ after all selection cuts.}
\vspace{-0.3cm}
\label{pic:minv2}
\end{figure}

\begin{figure}[htb]
\vspace{-1.1cm}
\centerline{
\includegraphics[clip,width=0.53\textwidth]{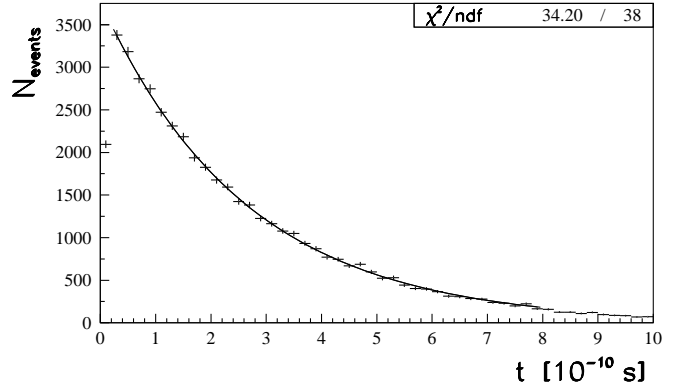}}
\vspace{-0.1cm}
\caption{Distribution of the reconstructed $\Lambda$ decay time
for measured events associated with the reaction \Lambdareaktionvorkomma. The solid line is explained in the text.}
%\vspace{0.4cm}
\label{fig:lebensdauerneu1}
\end{figure}

\begin{figure}[htb]
\vspace{-1.2cm}
\centerline{
\includegraphics[clip,width=0.53\textwidth]{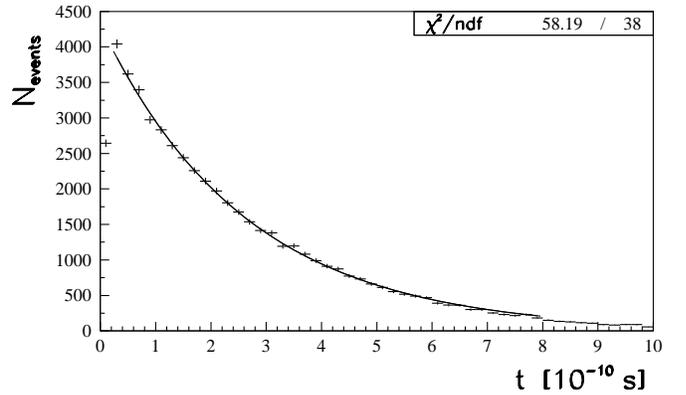}}
\vspace{-0.1cm}
\caption{Distribution of the reconstructed $\Lambda$ decay time for measured
events associated with the reaction \Sigmanreaktionvorkomma. The solid line is explained in the text.}
\label{fig:lebensdauerneu3}
\vspace{-0.2cm}
\end{figure}

The event samples were selected by a sequence of cuts. In a first
step events with one track of a negatively charged and two tracks of positively
charged particles were selected.
Then the reconstruction of a secondary vertex was required,
which could be a candidate for a \Lambdazerfall decay,
and the three-momentum of the $\Lambda$ candidates was reconstructed and
combined with the remaining positive track considered as a $K^+$ candidate
for the reconstruction of the vertex of the primary $\gamma\,p\,$ interaction.\\
In the next step of selection it was required that an event
passes a complete kinematic fit for the hypothesis \Lambdareaktionvorkomma\ or \Sigmanreaktion
at the primary vertex with \Lambdazerfall at the secondary vertex and,
in case of $\Sigma^0$, also with \Sigmazerfall as additional constraint.
If an event fitted both hypotheses, the reaction with the higher probability in the
kinematic fit was accepted.\\
For events which passed the fit procedure the missing mass was calculated from the
four-momenta (built up with the measured three-momenta combined with mass
assignments) of the particles at the primary vertex according to
\begin{displaymath}
  m_{\gamma\,p\,-\,K^+}\,=\,\sqrt{(p^{\,in}_\gamma\,+\,p^{\,in}_p\,-\,p^{\,out}_{K^+})^2}\,\,\,\,\,\,\,\,\,\,.
\end{displaymath}
The distribution of the missing mass is shown in Fig.~\ref{pic:minv0}.
Clear signals are seen for $\Lambda$ and $\Sigma^0$ as well as indications of excited
hyperons on top of background events.\\
The background was reduced by a sequence of cleaning cuts:
\begin{itemize}
  \item The events were required to have a primary vertex inside the sensitive target volume
(within resolution limits).
  \item The missing mass \miskaonvorkomma\ was required to be in
\linebreak[4]
the range $1000~\mbox{MeV}\,<\,~$\miskaon$\,<\,1240~\mbox{MeV}$ for events selected as \Lambdareaktionvorkomma\
and $1050~\mbox{MeV}\,<\,~$\miskaon$\,<\,1350~\mbox{MeV}$ for events selected as \Sigmanreaktionvorkomma.
  \item The invariant mass, reconstructed from the four-mo\-men\-ta of
the proton and pion at the secondary vertex (before the kinematic fit) according to\\
\begin{displaymath}
   m_\Lambda\,=\,\sqrt{\,(\,p^{\,out}_{\pi^-}+\,p^{\,out}_p\,)^2}\,\,\,\,\,,
   \label{eq:mvxfit}
\end{displaymath}
had to be consistent with the nominal $\Lambda$ mass within the range of $\pm 8$~MeV.
  \item The probability of the kinematic fits had to be greater than $10^{-10}$.
\end{itemize}
%probability of the kinematic fits:
%$P_{kin}(\chi^2)\,>\,10^{-10}$. 
%The effect of this cut is demonstrated by
%\begin{figure}[htb]
%\vspace{-1.5cm}
%\centerline{
%\includegraphics[clip,width=0.53\textwidth]{massev0iskaon1_paper.eps}}
%\vspace{-0.1cm}
%\caption{Distribution of the missing mass \miskaon of events that fitted at least one of the
%kinematic hypotheses \Lambdareaktionvorkomma\ and \Sigmanreaktionvorkomma\ with
%a probability better than $P_{kin}(\chi^2)\,>\,10^{-10}$.}
%\vspace{-0.1cm}
%\label{pic:minv1}
%\end{figure}
%Fig.~\ref{pic:minv1}.\\
After these cleaning cuts the data samples contained 51977 events selected as
\Lambdareaktionvorkomma\ and 54388 events selected as \Sigmanreaktionvorkomma. The missing
mass distributions of both event samples are shown in Fig.\ref{pic:minv2}. The
migration of events from one reaction to the other has been estimated by means
of Monte-Carlo simulated events and the cross sections were corrected correspondingly
\begin{figure}[htb]
\vspace{-1.3cm}
\centerline{
\includegraphics[clip,width=0.53\textwidth]{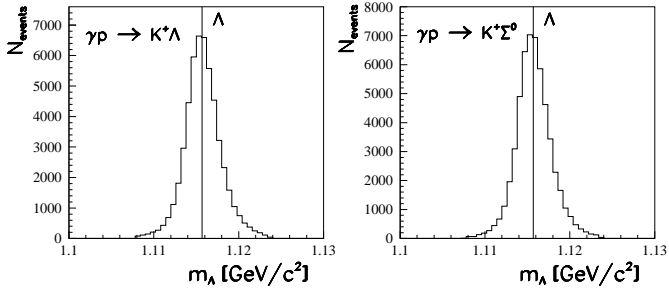}}
\vspace{-0.1cm}
\caption{Distribution of the invariant mass \mvxfit for the events selected as
\Lambdareaktionvorkomma\ and \Sigmanreaktionvorkomma\ respectively. The vertical
lines mark the nominal position of the $\Lambda$ mass.}
%\vspace{0.2cm}
\label{pic:minv3}
\end{figure}
\hspace{-0.15cm}(see Section~\ref{sec:mutual}).\\
Fig~\ref{pic:minv3} shows the distribution of the invariant mass \mvxfitvorkomma\
after the selection cuts.
The resolution of the reconstructed $\Lambda$ mass
is about 2~MeV.\\
%
%\section{Decay time distributions}
%
The reconstructed $\Lambda$ decay time distributions of both data sets ($K^+\Lambda, K^+\Sigma^0$)
are shown in Figures~\ref{fig:lebensdauerneu1} and~\ref{fig:lebensdauerneu3}.
The solid line describes a fit to an exponential
function $f(t)\,=\,a\,\cdot\,e^{-t/\tau_\Lambda}$
carried out in the time-range of constant acceptance without using the first
time bin. The parameter $a$ was determined
in the fits, while $\tau_\Lambda$ has been fixed to the PDG
\hspace{-0.15cm}lifetime value \cite{pdg}. The decay time distributions are well described by the fits for
both reactions (the values of $\chi^2$ per degrees of
freedom, $ndf$, are given in the figures).\\
Possible background from other reactions which have two positively charged and one negatively
charged particles in the final state, such as \Zweipireaktion\ and
\Dreipireaktion\ (\Neutrondreipivorkomma\ is suppressed by the kinematic fit),
are expected to contribute mainly at small decay times. A corresponding
excess of events is not visible. Instead, a loss of events is observed in the first
$t$ bin ($0\,<\,t\,<\,10^{-10}$~s), which is due to a migration effect caused by the limited resolution
of the reconstructed positions of primary and secondary vertices, which was verified
with Monte-Carlo simulated events \cite{glander}.

\section{Acceptance corrections}
\label{sec:acceptance}

\begin{figure}[htb]
\vspace{-0.7cm}
\centerline{
\includegraphics[clip,width=0.55\textwidth]{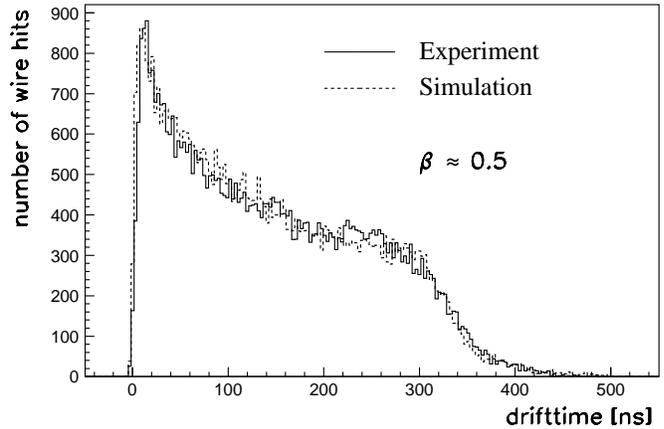}}
\vspace{-0.1cm}
\caption{Drift time spectrum for drift times produced by particles in experimental and simulated events of
reaction \Zweipireaktion\ with velocities $\beta~\approx~0.5$.}
\label{fig:drifttimes1}
\vspace{-0.8cm}
\end{figure}

\begin{figure}[htb]
\vspace{-0.2cm}
\centerline{
\includegraphics[clip,width=0.55\textwidth]{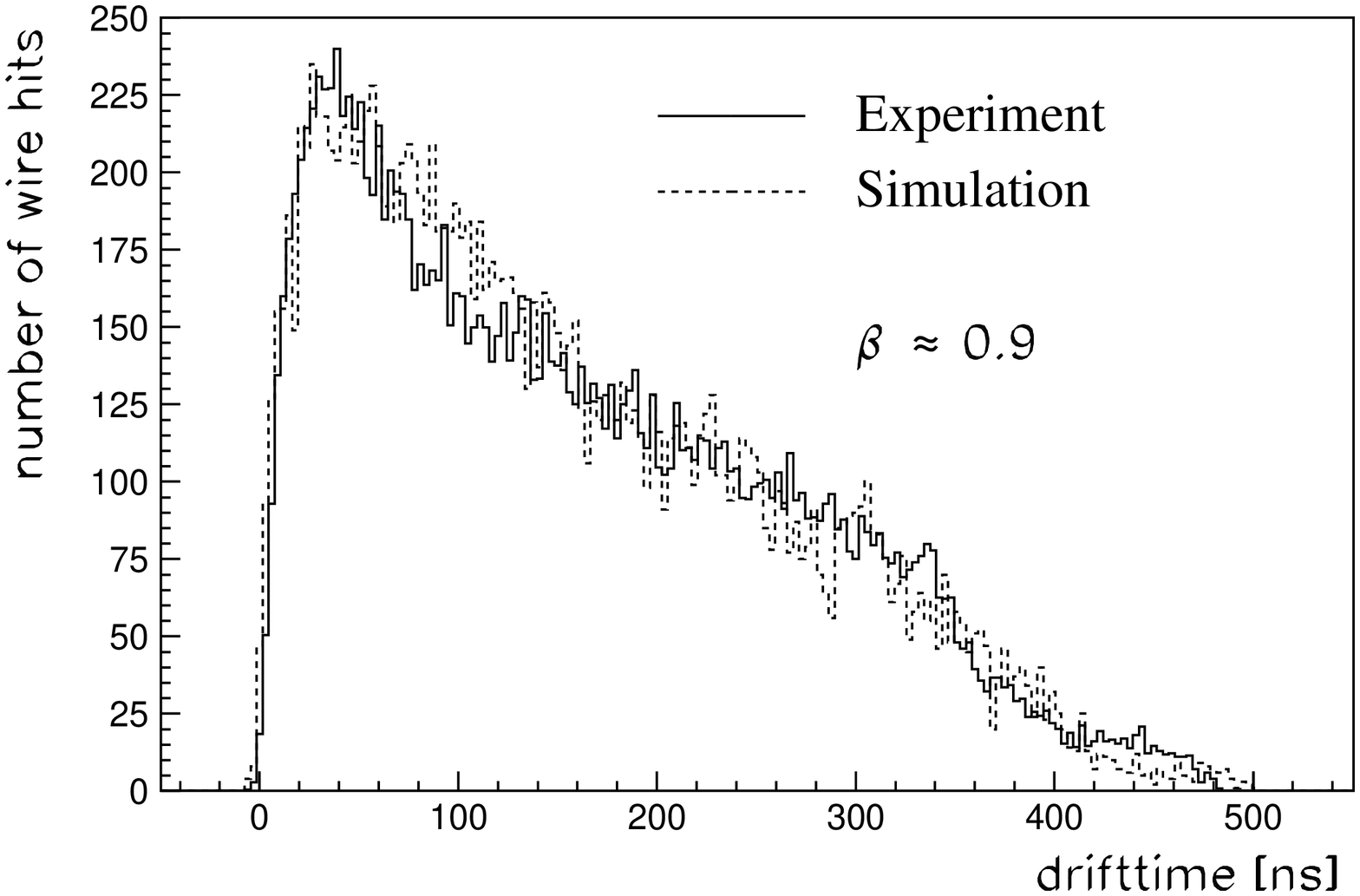}}
\vspace{-0.1cm}
\caption{Drift time spectrum for drift times produced by particles in experimental and simulated events of
reaction \Zweipireaktion\ with velocities $\beta~\approx~0.9$.}
\label{fig:drifttimes2}
\vspace{-0.4cm}
\end{figure}

The acceptance of events, which were measured in the SAPHIR setup and had passed reconstruction
and selection, was calculated using a simulation package which was developed for SAPHIR based
on GEANT 2.0 \cite{geant}.\\
The simulation of the response of the detector components comprised the photon tagging system
\cite{omega,hannappeld,barth2}, the scintillator hodoscopes \cite{neuerburgd,barth2}
and the drift chamber system \cite{glander}.
The efficiencies of the tagging system and scintillator hodoscopes, which
together with the beam veto count\-er defined the trigger efficiency of the data taking, were determined
separatly from the data of the four data taking periods and included in the simulations.\\
The simulation of the drift chamber system was carried out at a very detailed level \cite{glander}.
In particular, the statistically distribution of the ionization deposits
along a track of a charged particle in a drift cell were taken into account
so that the drift length of the electron cluster, which is closest
to the signal wires, can be longer than
\begin{figure}[htb]
%\vspace{-0.2cm}
\centerline{
\includegraphics[clip,width=0.43\textwidth]{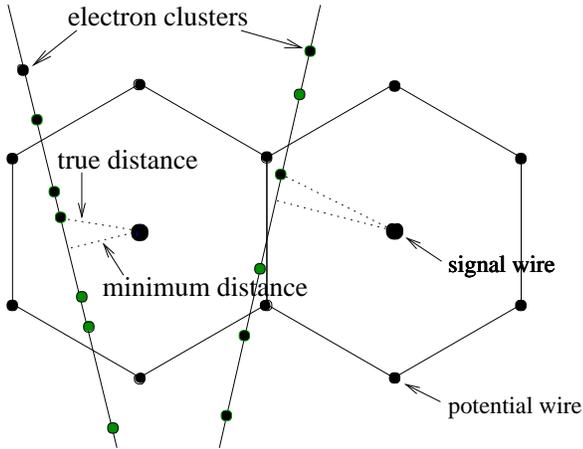}}
%\vspace{-0.2cm}
\caption{Two particles, passing the hexagonal drift cells of a planar drift chamber, ionizising
drift chamber gas molecules as germs for the production of electron clusters along the
particle tracks.}
\label{fig:primaerionisation}
\vspace{-0.1cm}
\end{figure}
the distance of closest approach of the track to the signal wire (see\ Fig.~\ref{fig:primaerionisation}).
With the gas mixture used for the chambers (see Section~\ref{sec:datataking}) this effect
was important. While the mean free path length between two
ionization deposits, for a particle velocity $\beta\,\approx\,0.5$, is of the order of 200~$\mu$m,
which corresponds to the spatial resolution of the chambers\footnote{The spatial resolution is a
function of $\beta$ and changes between 150~$\mu$m and 300~$\mu$m \cite{glander}.},
for $\beta\,\approx\,0.9$ it increases up to 2~mm \cite{glander}.\\
Drift times simulated at this level describe measured distributions well. This is demonstrated
for the final state $p\,\pi^+\,\pi^-$ and two $\beta$ values in Figs.~\ref{fig:drifttimes1}
and~\ref{fig:drifttimes2}.
If the ionization cloud, which determines the drift time, is displaced from the point of closest
approach, the actually measured drift time is delayed and the drift distance calculated from the
drift time appears prolongated with respect to the minimum distance.
Because the mean free path length grows with $\beta$ one observes on average a shift of drift
times to higher values with increasing $\beta$ values
(see Figs.~\ref{fig:drifttimes1} and~\ref{fig:drifttimes2}).
This is most obvious when looking at the rise of both spectra, which is
steeper for low $\beta$, and the broadening of the maximum with $\beta$.\\
The spread of the ionization clouds along the track according to the free path length in the chamber
gas also affected the reconstruction of the tracks, because hits with significantly prolongated drift
distances disturb the track fit calculations directly.
That means: In case that the difference between the true and the minimum distance\
(see Fig.~\ref{fig:primaerionisation}) is much larger than the drift chamber resolution,
such hits give large contributions to the $\chi^2$ value of the track fit, because in the track
reconstruction it has to be assumed that drift times are related to the point of closest approach.
Such disturbing hits were removed after the first track fit iteration:
If the determined drift distance differed by more than three times the maximal
spatial resolution of 300~$\mu$m from the distance between the fitted track and the signal wire
after the first fit, then the hit was rejected.
\begin{figure}[htb]
\vspace{-0.8cm}
\centerline{
\includegraphics[clip,width=0.55\textwidth]{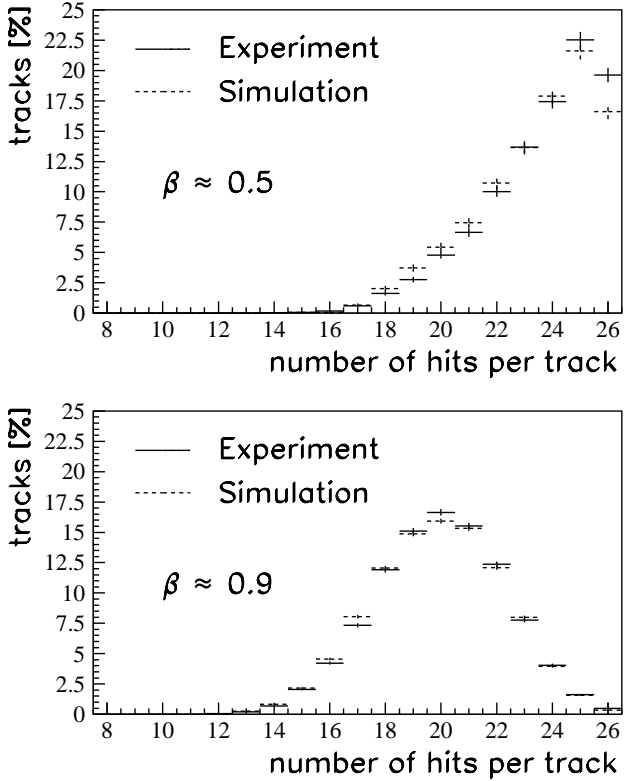}}
\vspace{-0.2cm}
\caption{Number of hits per track after removing bad hits from the track fit.}
\label{fig:hits}
\vspace{-0.1cm}
\end{figure}
\hspace{-0.15cm}This procedure removed also hits with other problems in the track reconstruction like wrongly
resolved left-right ambiguities between tracks and signal wires,
to give just an example.\\
The final track fit was carried out with the remaining hits in the central drift
chamber (14 detection layers) and in the forward chamber (12 layers).
The distribution of the number of
hits used for the final track fit is well described by simulated events. This is
shown in Fig.~\ref{fig:hits}, again for $\beta\,\approx\,0.5$ and $\beta\,\approx\,0.9$.\\
The acceptance was calculated from simulated events as the ratio of finally accepted
to generated events in bins of photon energy and kaon production angle in the {\it cms}.
The acceptance values for both reactions, \Lambdareaktionvorkomma\ and \Sigmanreaktionvorkomma,
\begin{figure}[htb]
%\vspace{-0.2cm}
\centerline{
\includegraphics[clip,width=8.6cm]{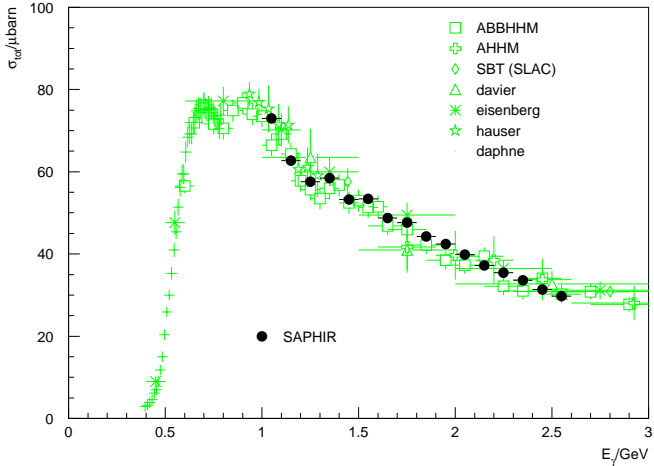}}
\vspace{-0.2cm}
\caption{The total cross section for the reaction \Zweipireaktion\ measured by SAPHIR \cite{barth2}
in comparison with existing world data \cite{landolt}.}
\label{fig:2piwqtot}
\vspace{-0.1cm}
\end{figure}
are of the order of 10\% and do nowhere vanish throughout the full kinematic range.\\
The same simulation package was used for other reactions measured at SAPHIR \cite{barth2,omega,phi,rho}.
The cross sections are in general consistent with existing world data in normalization
as well as in energy dependence. As an example, Fig.~\ref{fig:2piwqtot} shows this agreement for the reaction
cross section of \Zweipireaktion\ which has been measured at SAPHIR with high statistics \cite{barth2}.
This verifies that the acceptance calculations are well understood.

\section{Cross sections}
\label{sec:cross}

\begin{figure*}[htb]
\vspace{0.4cm}
\centerline{
\includegraphics[clip,height=1.30\textwidth,width=1.05\textwidth]{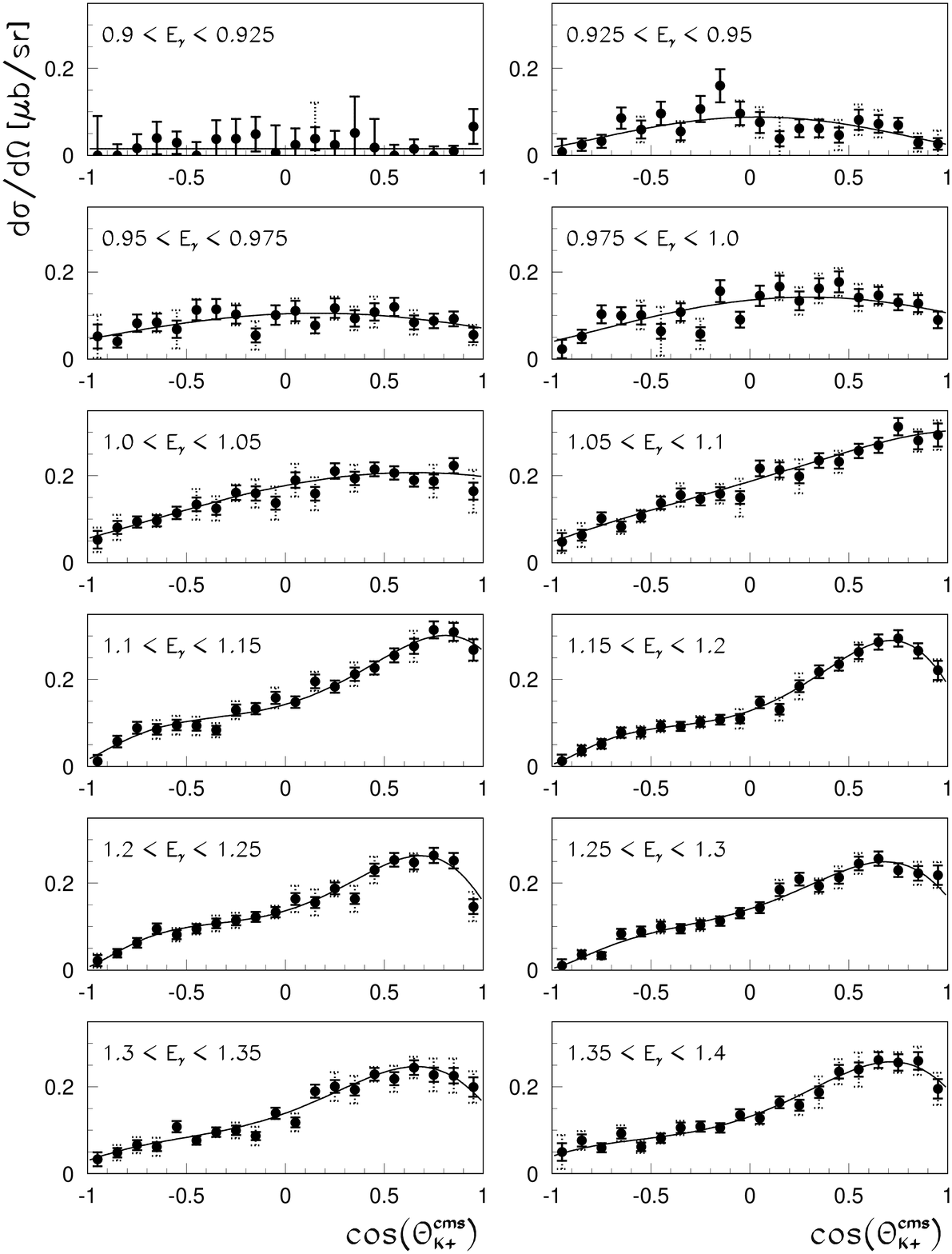}}
\vspace{-0.2cm}
\caption{Differential cross sections of \Lambdareaktionvorkomma\ for photon energies $0.9~\mbox{GeV}\,<\,E_\gamma\,<\,1.4~\mbox{GeV}$.
The solid and dashed bars represent errors $\sigma_w$ and $\sigma_d$ (see text).
The solid lines describe fits of Legendre polynomials to the data.}
\label{fig:wqdifflambda1}
\end{figure*}

\begin{figure*}[htb]
\vspace{0.4cm}
\centerline{
\includegraphics[clip,height=1.30\textwidth,width=1.05\textwidth]{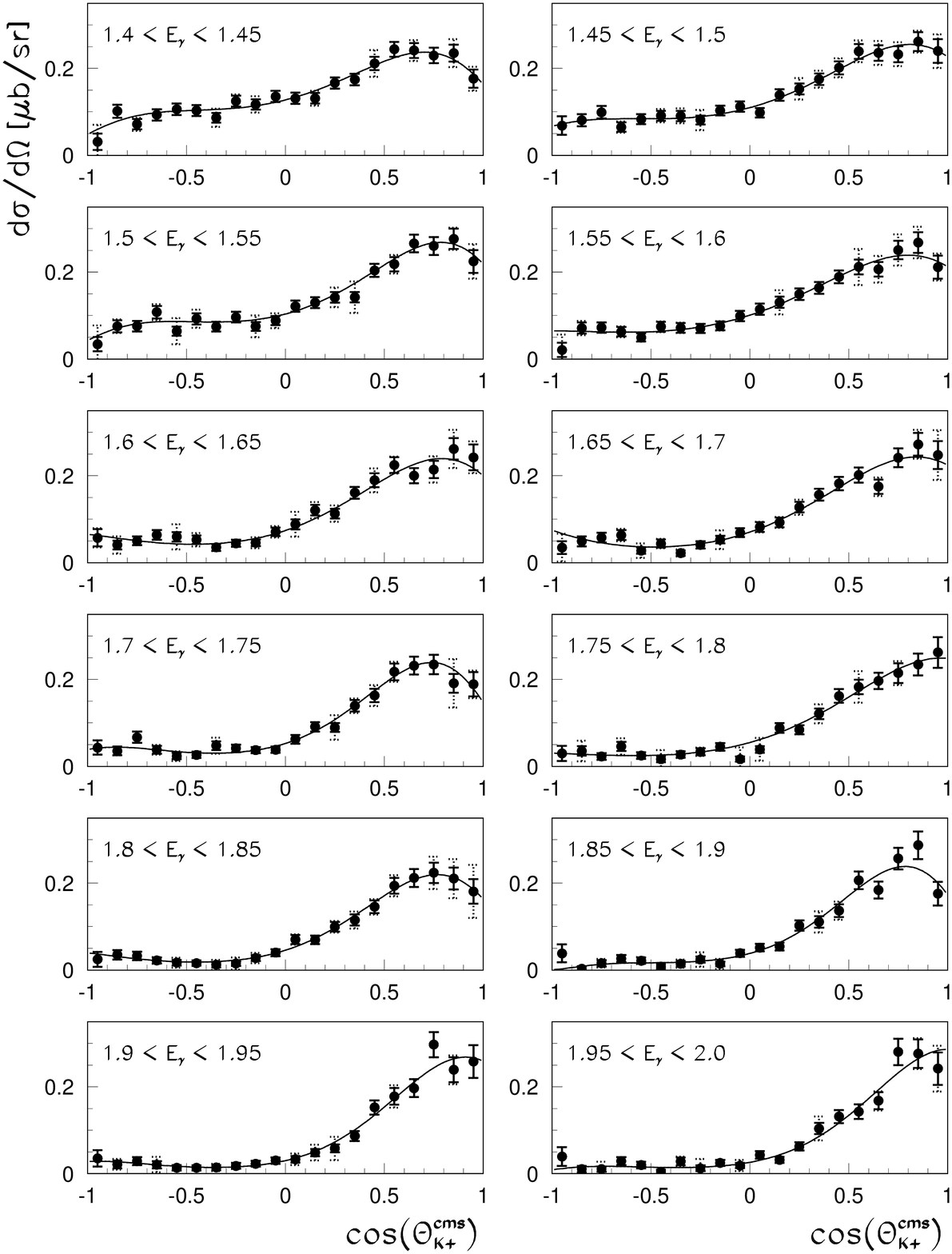}}
\vspace{-0.2cm}
\caption{Differential cross sections of \Lambdareaktionvorkomma\ for photon energies $1.4~\mbox{GeV}\,<\,E_\gamma\,<\,2.0~\mbox{GeV}$.
The solid and dashed bars represent errors $\sigma_w$ and $\sigma_d$ (see text).
The solid lines describe fits of Legendre polynomials to the data.}
\label{fig:wqdifflambda2}
\end{figure*}

\begin{figure*}[htb]
\vspace{0.4cm}
\centerline{
\includegraphics[clip,height=1.30\textwidth,width=1.05\textwidth]{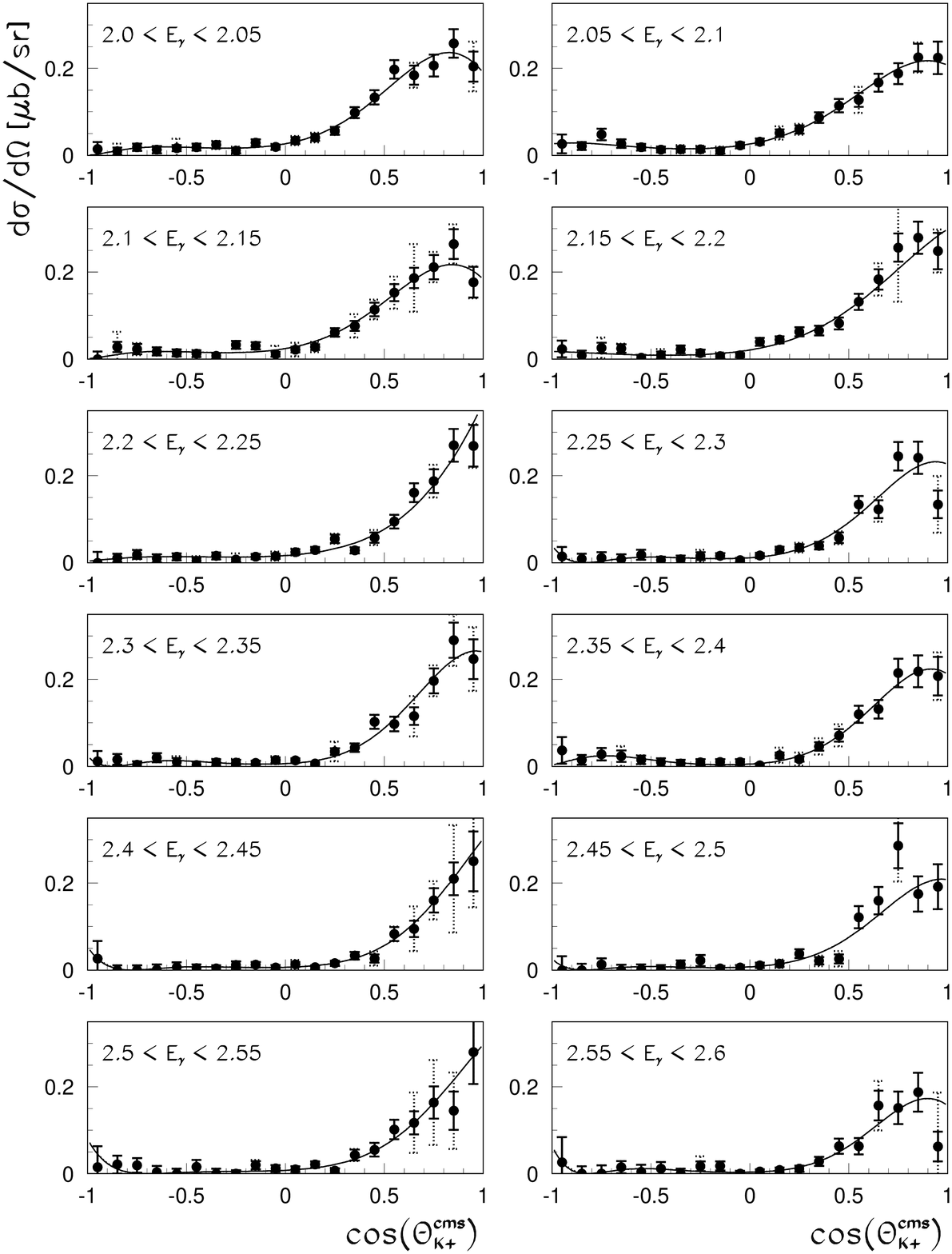}}
\vspace{-0.2cm}
\caption{Differential cross sections of \Lambdareaktionvorkomma\ for photon energies $2.0~\mbox{GeV}\,<\,E_\gamma\,<\,2.6~\mbox{GeV}$.
The solid and dashed bars represent errors $\sigma_w$ and $\sigma_d$ (see text).
The solid lines describe fits of Legendre polynomials to the data.}
\label{fig:wqdifflambda3}
\end{figure*}

\begin{figure*}[htb]
%\vspace{-0.2cm}
\centerline{
\includegraphics[clip,width=0.94\textwidth]{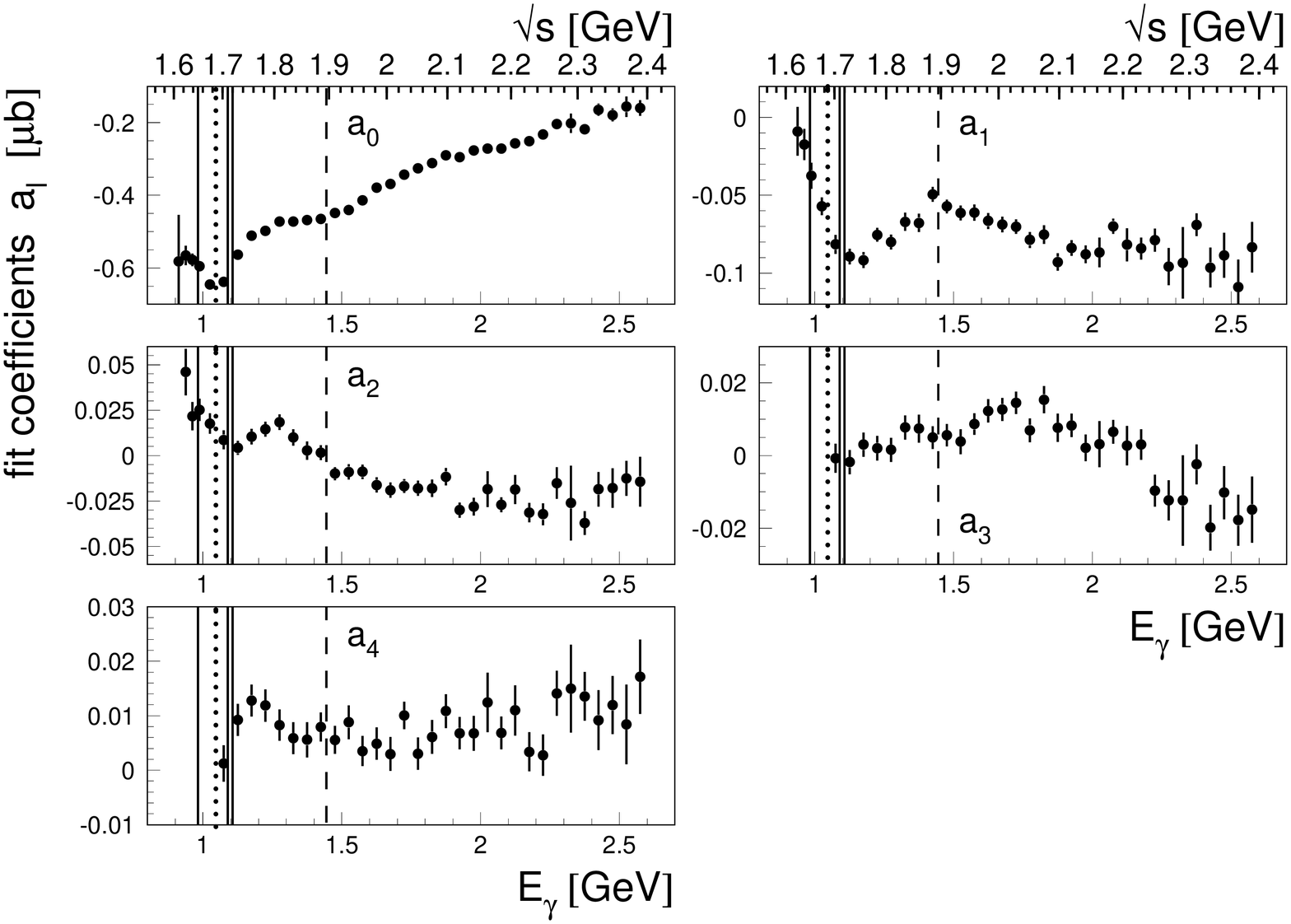}}
\vspace{-0.1cm}
\caption{Legendre polynomial fit coefficients from differential cross sections for the reaction
\Lambdareaktionvorkomma\ as a function of the photon energy. For explanation of the vertical lines
see caption of Fig.~\ref{fig:wqtotlambda}.}
\label{fig:koeffklambda}
\end{figure*}

\begin{figure*}[htb]
\vspace{0.2cm}
\centerline{
\includegraphics[clip,width=0.94\textwidth]{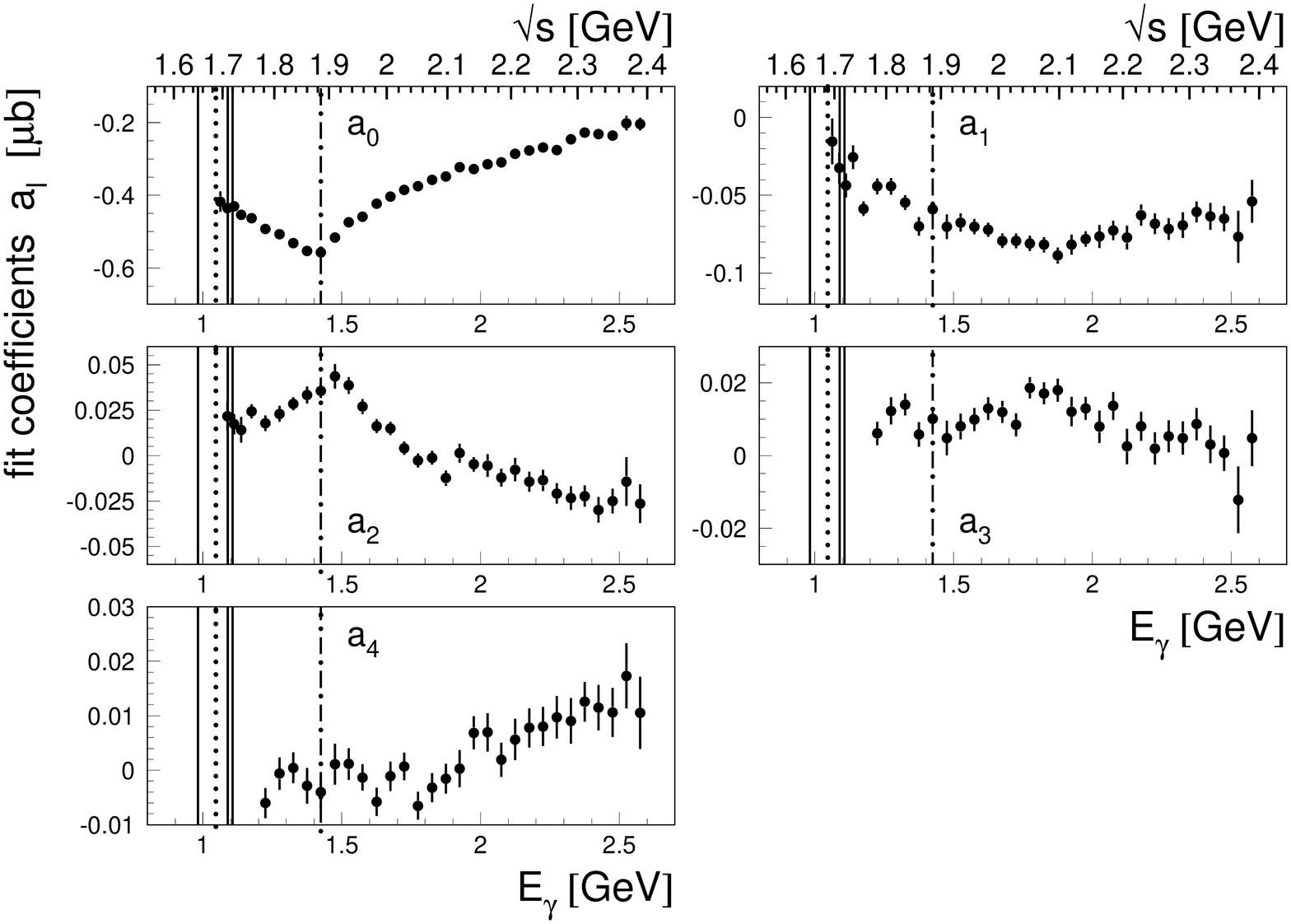}}
\vspace{-0.1cm}
\caption{Legendre polynomial fit coefficients from differential cross sections for the reaction
\Sigmanreaktionvorkomma\ as a function of the photon energy. For explanation of the vertical lines
see caption of Fig.~\ref{fig:wqtotsigma}.}
\label{fig:koeffksigma0}
\vspace{-0.5cm}
\end{figure*}

Cross sections $x_i$ and their uncertainties $\sigma_i$ ($i\,=\,1,\ldots,N$ where $N$ is the number of
data taking periods; $N$ is equal 4 or 3 respectively\footnote{Three
of the four data taking periods were carried out with an energy of the primary electron beam
of $E_0\,=\,2.8~\mbox{GeV}$. During one run
an electron beam with an energy of $E_0\,=\,2.6~\mbox{GeV}$ was used instead.
Therefore, for the determination of the differential cross sections the data
of this run were not used for photon energies, $E_\gamma$, above 2.4~GeV.})
were determined as a function of photon energy and $K^+$ production
angle in the overall {\it cms} separately for the data taking periods during the years
1997/98.
The statistically weighted mean $m$ of the four (three) measurements $x_i$ and its statistical
error $\sigma_w$ were determined according to
\begin{displaymath}
  m\,=\,\frac{\sum\limits_{i\,=\,1}^{N}\,\frac{x_i}{\sigma_i^2}}{\sum\limits_{i\,=\,1}^{N}\,\frac{1}{\sigma_i^2}}\,\,\,\,\,\,\,\,\,\,\,\,\,\,,\,\,\,\,\,\,\,\,\,\,\,\,\,\,\sigma_w\,=\,\frac{1}{\sqrt{\sum\limits_{i\,=\,1}^{N}\,\frac{1}{\sigma_i^2}}}\,\,\,\,\,\,\,\,\,\,\,\,\,\,.
\end{displaymath}
The results on differential cross sections as a function of the $K^+$ production
angle in the overall {\it cms} in bins of photon energy are shown for \Lambdareaktion
in Figures~\ref{fig:wqdifflambda1},~\ref{fig:wqdifflambda2} and~\ref{fig:wqdifflambda3}
and for \Sigmanreaktionvorkomma\ in
Figures~\ref{fig:wqdiffsigma1},~\ref{fig:wqdiffsigma2} and~\ref{fig:wqdiffsigma3},
and the numbers are given in Appendix~\ref{app:diff}.
The data points in Figures~\ref{fig:wqdifflambda1} to~\ref{fig:wqdiffsigma3} are the
determined weighted means $m$. The solid bars represent the statistical errors $\sigma_w$.\\
These errors, $\sigma_w$, are based on the full statistics of the event samples, but cannot
account for systematic uncertainties caused by differences in the run conditions.
Known differences were taken into account in the simulation used for acceptance corrections.
In order to take into account possible additional fluctuations and uncertainties between the
runs, which were not explicitly considered in the simulations,
another error $\sigma_d$ has been calculated which is the standard deviation
of the four measurements $x_i$ to the weighted mean $m$
given by the square root of the variance $V_m$:\\
\begin{displaymath}
  \sigma_d\,=\,\sqrt{V_m}\,=\,\sqrt{\frac{1}{N\,(N\,-\,1)}\,\sum_{i\,=\,1}^{N}\,(m\,-\,x_i)^2}\,\,\,\,\,\,\,\,\,\,\,\,.
\end{displaymath}
The dashed bars in Figures~\ref{fig:wqdifflambda1} to~\ref{fig:wqdiffsigma3} show the errors $\sigma_d$
obtained for the $N=4$ runs. In general $\sigma_d$ is bigger
than $\sigma_w$. In cases where $\sigma_d$ was smaller than $\sigma_w$,
$\sigma_d$ was set equal to $\sigma_w$. $\sigma_w$ and $\sigma_d$ were determined
as a function of photon energy and kaon production angle. The error defined by
$\sqrt{{\sigma_d}^2-{\sigma_w}^2}$ is a measure of the systematic uncertainties due to
fluctuations in the run conditions.\\
The solid lines in the Figs.~\ref{fig:wqdifflambda1}-\ref{fig:wqdiffsigma3} describe fits with
Legendre polynomials of the form
\begin{displaymath}
\frac{d\sigma}{d\,cos(\theta^{\,cms}_{K^+})} \,=\, \frac{q}{k} \, \left(\,\sum_{l\,=\,0}^{L\,=\,4} \, {a_l} \, P_l(cos \, \theta^{\,cms}_{K^+})\right)^2\,\,\,\,\,\,\,\,\,\,\,\,.
\end{displaymath}
The phase space factor $q/k$ is given by the ratio of kaon to photon momentum in the
overall {\it cms}.\\
The fitted coefficients $a_l$ of the Legendre polynomials are shown as a function of
photon energy in Figures~\ref{fig:koeffklambda} and~\ref{fig:koeffksigma0}. The coefficients show clear
structures indicating rapid
\linebreak[4]
changes of the production mechanisms.\\
It is worth to mention that the energy dependence of the coefficients $a_0$, $a_1$ and $a_2$
in Figures~\ref{fig:koeffklambda} and~\ref{fig:koeffksigma0}
depends only weakly on the choice of the maximum angular
momentum, $L$, and good $\chi^2$ values for the Legendre fits were obtained for
$L$-values between 2 and 4.
For
\linebreak[4]
$L\,=\,4$ the mean of all $\chi^2$ values is 0.8 and 90\% of all fits 

\clearpage
\begin{figure*}[htb]
\vspace{0.4cm}
\centerline{
\includegraphics[clip,height=1.30\textwidth,width=1.05\textwidth]{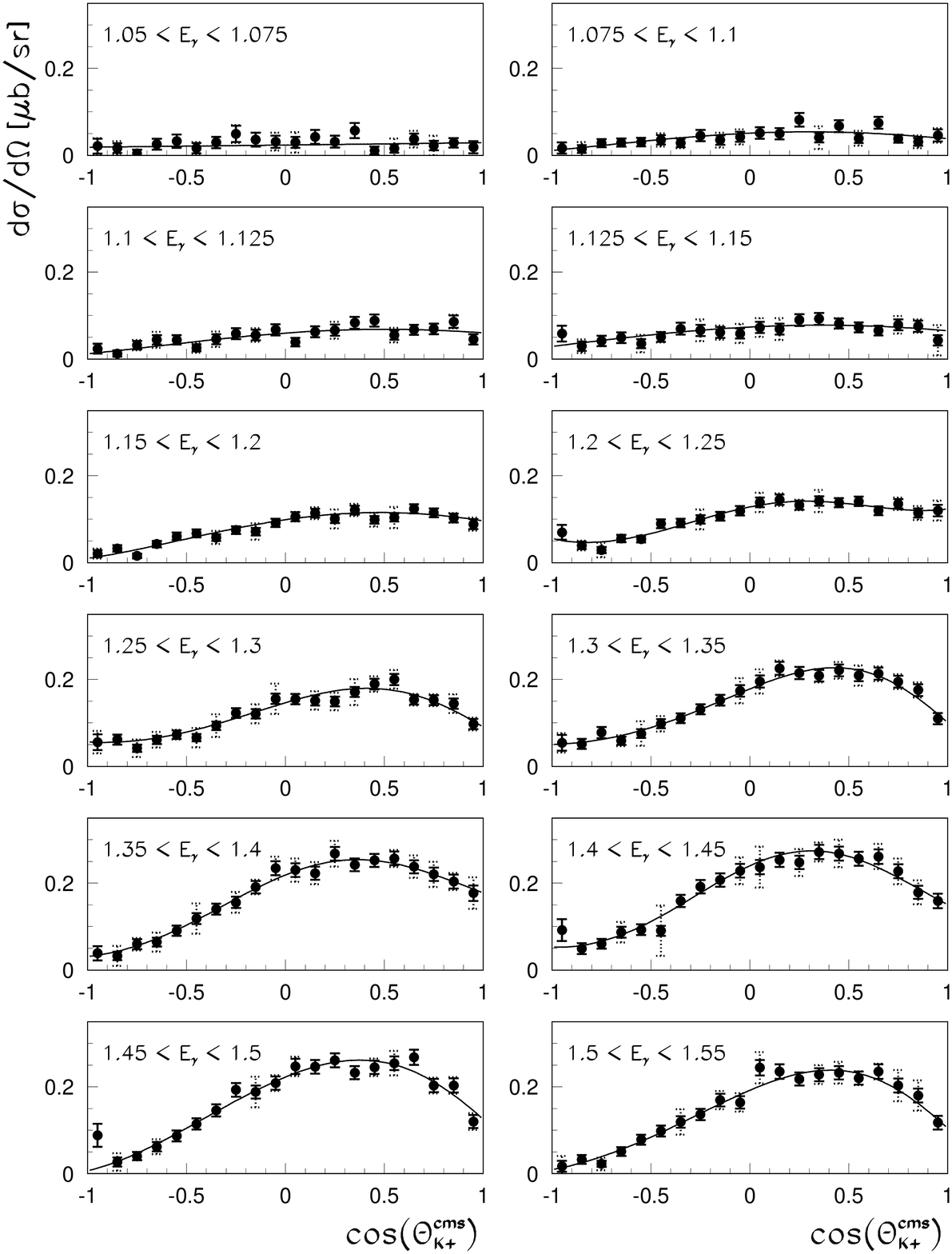}}
\vspace{-0.2cm}
\caption{Differential cross sections of \Sigmanreaktionvorkomma\ for photon energies $1.05~\mbox{GeV}\,<\,E_\gamma\,<\,1.55~\mbox{GeV}$.
The solid and dashed bars represent errors $\sigma_w$ and $\sigma_d$ (see text).
The solid lines describe fits of Legendre polynomials to the data.}
\label{fig:wqdiffsigma1}
\end{figure*}
\begin{figure*}[htb]
\vspace{0.4cm}
\centerline{
\includegraphics[clip,height=1.30\textwidth,width=1.05\textwidth]{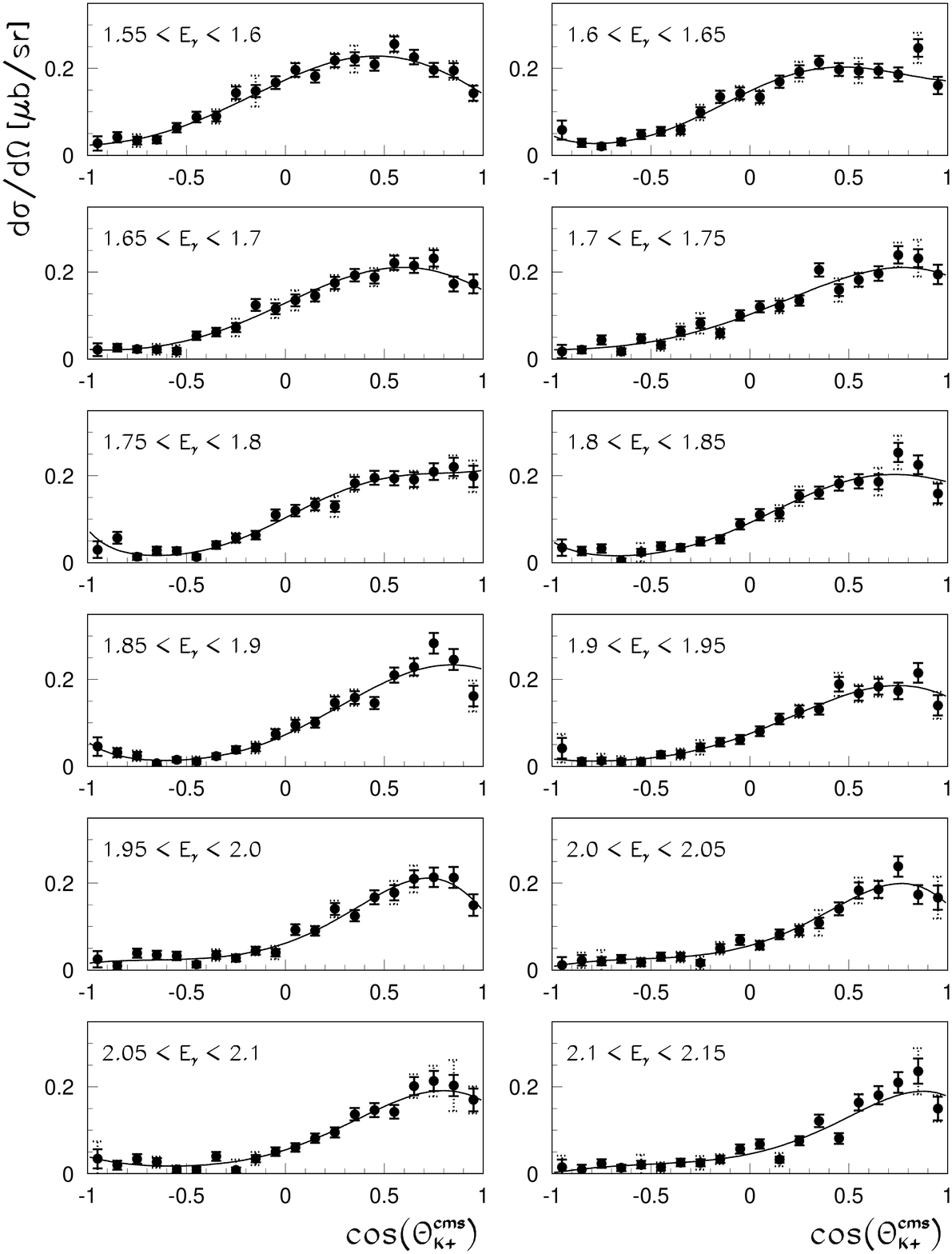}}
\vspace{-0.2cm}
\caption{Differential cross sections of \Sigmanreaktionvorkomma\ for photon energies $1.55~\mbox{GeV}\,<\,E_\gamma\,<\,2.15~\mbox{GeV}$.
The solid and dashed bars represent errors $\sigma_w$ and $\sigma_d$ (see text).
The solid lines describe fits of Legendre polynomials to the data.}
\label{fig:wqdiffsigma2}
\end{figure*}
\begin{figure*}[htb]
\vspace{1.8cm}
\centerline{
\includegraphics[clip,height=1.30\textwidth,width=1.05\textwidth]{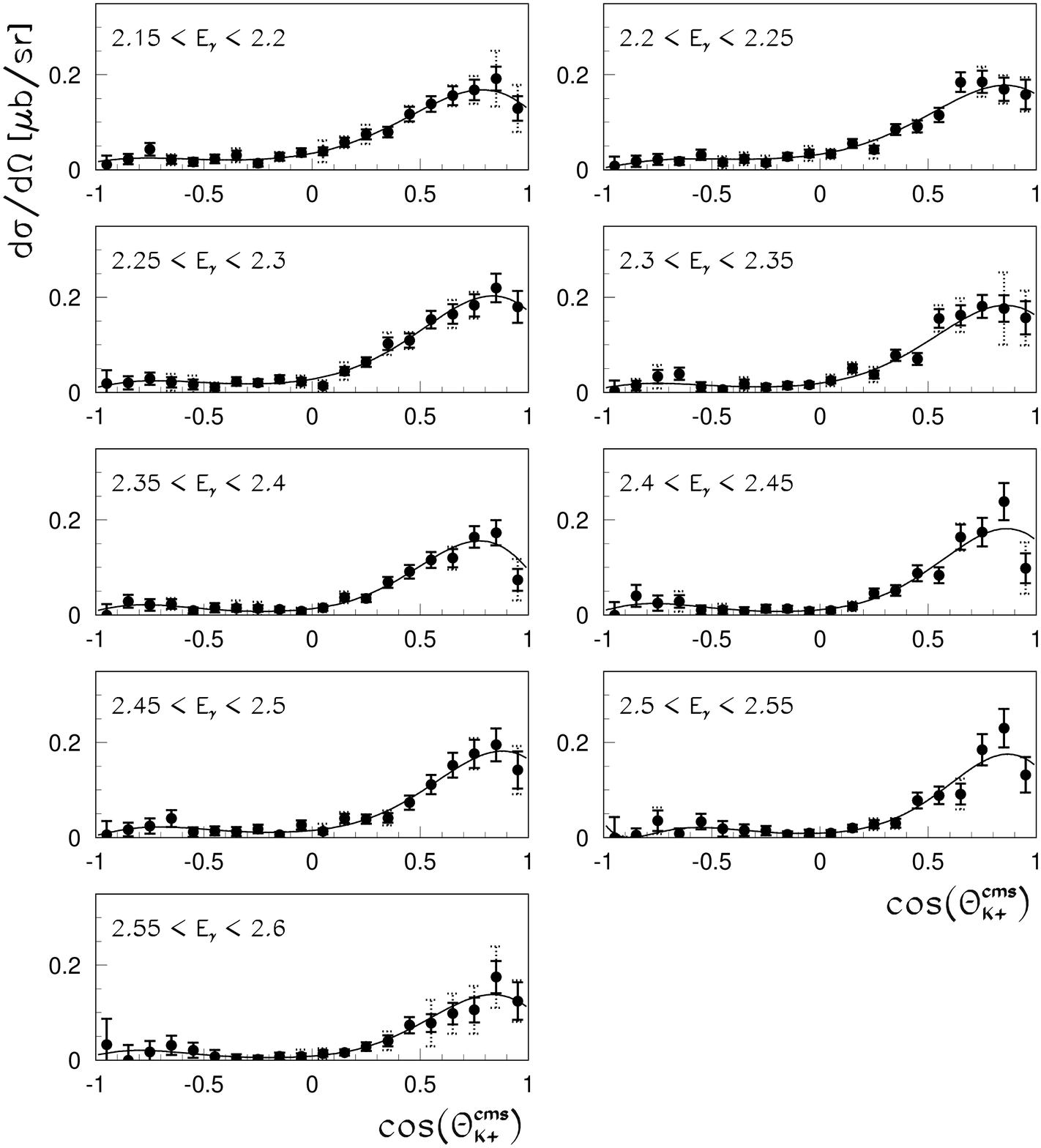}}
\vspace{-3.5cm}
\caption{Differential cross sections of \Sigmanreaktionvorkomma\ for photon energies $2.15~\mbox{GeV}\,<\,E_\gamma\,<\,2.6~\mbox{GeV}$.
The solid and dashed bars represent errors $\sigma_w$ and $\sigma_d$ (see text).
The solid lines describe fits of Legendre polynomials to the data.}
\label{fig:wqdiffsigma3}
\vspace{1.5cm}
\end{figure*}
\clearpage
\clearpage
%\begin{figure*}[htb]
%%\vspace{-0.2cm}
%\centerline{
%\includegraphics[clip,width=0.94\textwidth]{wqfitkoeff_klambda_cospolynom.eps}}
%\vspace{-0.1cm}
%\caption{Cosinus polynomial fit coefficients from differential cross sections for the reaction
%\Lambdareaktionvorkomma\ as a function of the photon energy. For explanation of the drawn in lines
%see caption of Fig.~\ref{fig:wqtotlambda}.}
%\end{figure*}
%\begin{figure*}[htb]
%\vspace{0.2cm}
%\centerline{
%\includegraphics[clip,width=0.94\textwidth]{wqfitkoeff_ksigma0_cospolynom.eps}}
%\vspace{-0.1cm}
%\caption{Cosinus polynomial fit coefficients from differential cross sections for the reaction
%\Sigmanreaktionvorkomma\ as a function of the photon energy. For explanation of the drawn in lines
%see caption of Fig.~\ref{fig:wqtotsigma}.}
%\vspace{-0.5cm}
%\end{figure*}
%\clearpage
\begin{figure*}[htb]
\centerline{
\includegraphics[clip,width=1.04\textwidth]{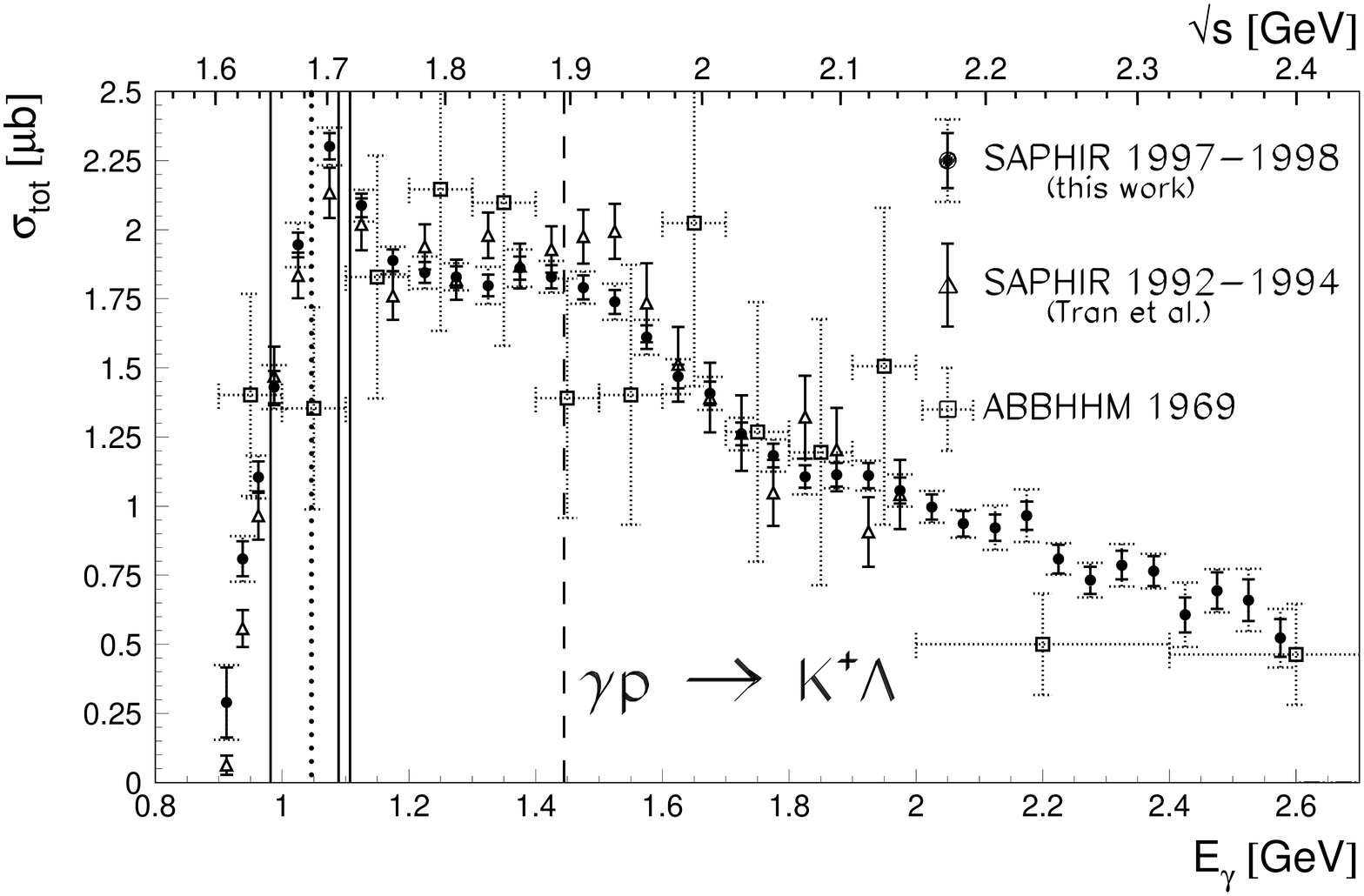}}
%\vspace{-0.1cm}
\caption{Total cross section of \Lambdareaktionvorkomma.
%The estimated background from reaction \Sigmanreaktionvorkomma\ is shown as hatched area.
The vertical lines indicate the threshold energy of \Sigmanreaktionvorkomma\ (dotted),
mass values of the known resonances $S_{11}(1650)$, $P_{11}(1710)$ and
$P_{13}(1720)$ (solid) and the position of the hypothetical resonance $D_{13}(1895)$ (dashed)
discussed in the text.}
\label{fig:wqtotlambda}
\end{figure*}
\begin{figure*}[htb]
\vspace{-0.1cm}
\centerline{
\includegraphics[clip,width=1.04\textwidth]{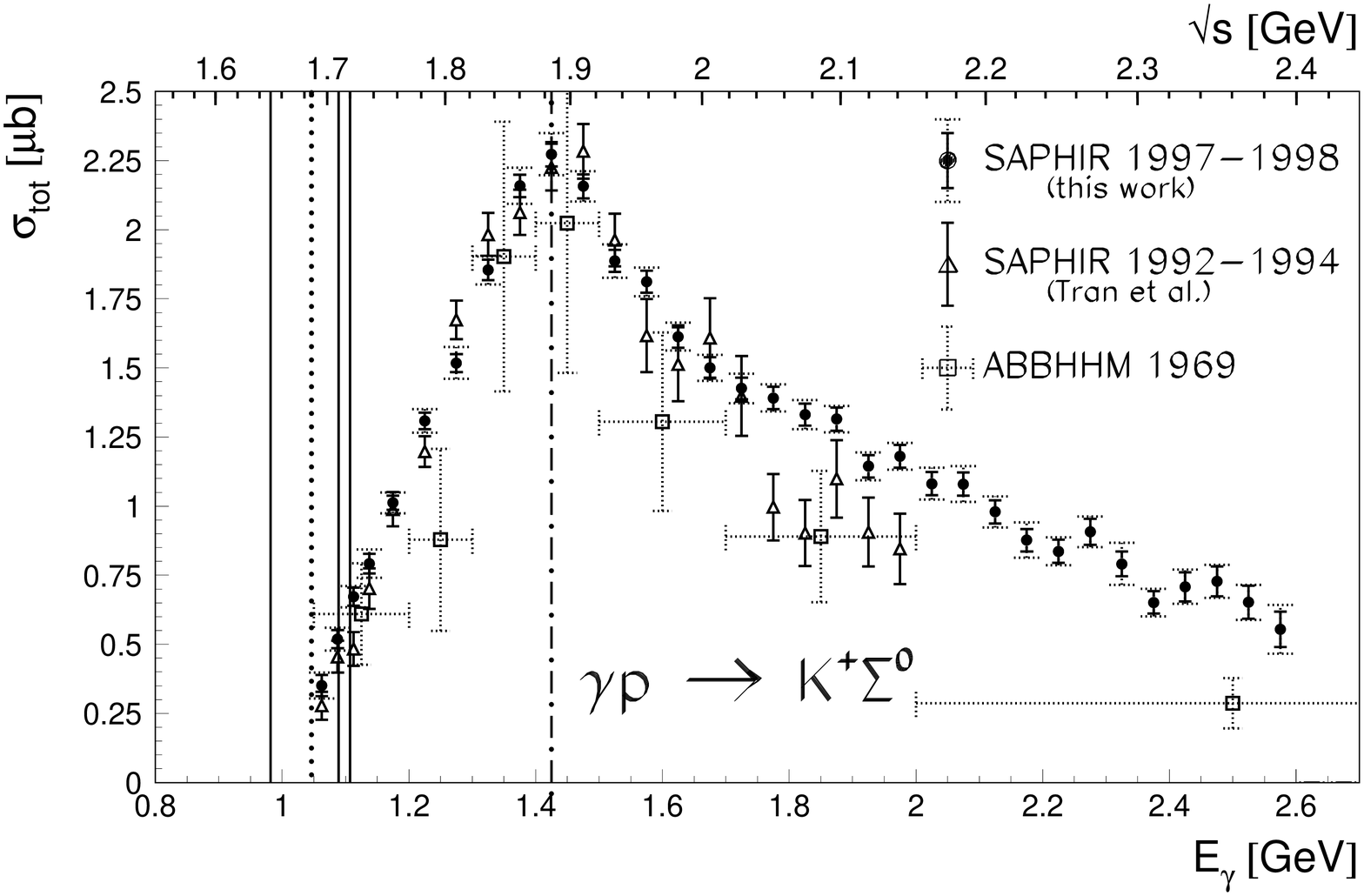}}
%\vspace{-0.1cm}
\caption{Total cross section of \Sigmanreaktionvorkomma.
%The estimated background from reaction \Lambdareaktionvorkomma\ is shown as hatched area.
The vertical lines indicate the threshold energy of \Sigmanreaktionvorkomma\ (dotted) and the
mass values of the known resonances $S_{11}(1650)$, $P_{11}(1710)$ and
$P_{13}(1720)$ (solid). The maximum cross section (dash-dotted line) is discussed with respect
to contributions of $\Delta$-resonances at this energy region (see text).}
\label{fig:wqtotsigma}
\vspace{-1.0cm}
\end{figure*}

\clearpage

\hspace{-0.53cm}to \Lambdareaktionvorkomma\ and \Sigmanreaktionvorkomma\
give values for $\chi^2$ between 0.4 and 1.4. For $L\,=\,3$ the
mean of $\chi^2$ is 1.1 and 90\% of the $\chi^2$ values vary between 0.5 and 1.7. For $L\,=\,2$
the mean is 1.3 with 90\% of the $\chi^2$ between 0.6 and 2.3. Because the fits for $L\,=\,4$ deliver
the best description over the full angular range (especially for the very forward and backward regions),
$L\,=\,4$ was used for the fits shown in Figures~\ref{fig:wqdifflambda1}-\ref{fig:wqdiffsigma3}.\\
The reaction cross sections were determined as a function of energy
by integrating the differential cross section over the angular range. They are
shown for \Lambdareaktionvorkomma\ in Fig.~\ref{fig:wqtotlambda} and for
\Sigmanreaktionvorkomma\ in Fig.~\ref{fig:wqtotsigma}.\\
Solid and dashed error bars were determined for the total cross sections by quadratic addition
of the single errors in the angular bins of the differential cross sections per energy bin,
separately for $\sigma_w$ and $\sigma_d$.\\
For comparison, previous measurements of the total cross sections are also shown in
Figures~\ref{fig:wqtotlambda} and~\ref{fig:wqtotsigma}:
the measurements of the ABBHHM collaboration~\cite{ABBHHM} and the SAPHIR cross section from
the first data taking periods in 1992-1994 \cite{tran} (this older SAPHIR data are not included
in the current analysis).

\section{Mutual event migrations of reactions \Lambdareaktionvorkomma\ and \Sigmanreaktion}
\label{sec:mutual}

\begin{figure}[htb]
\vspace{-1.5cm}
\centerline{
\includegraphics[clip,width=0.45\textwidth]{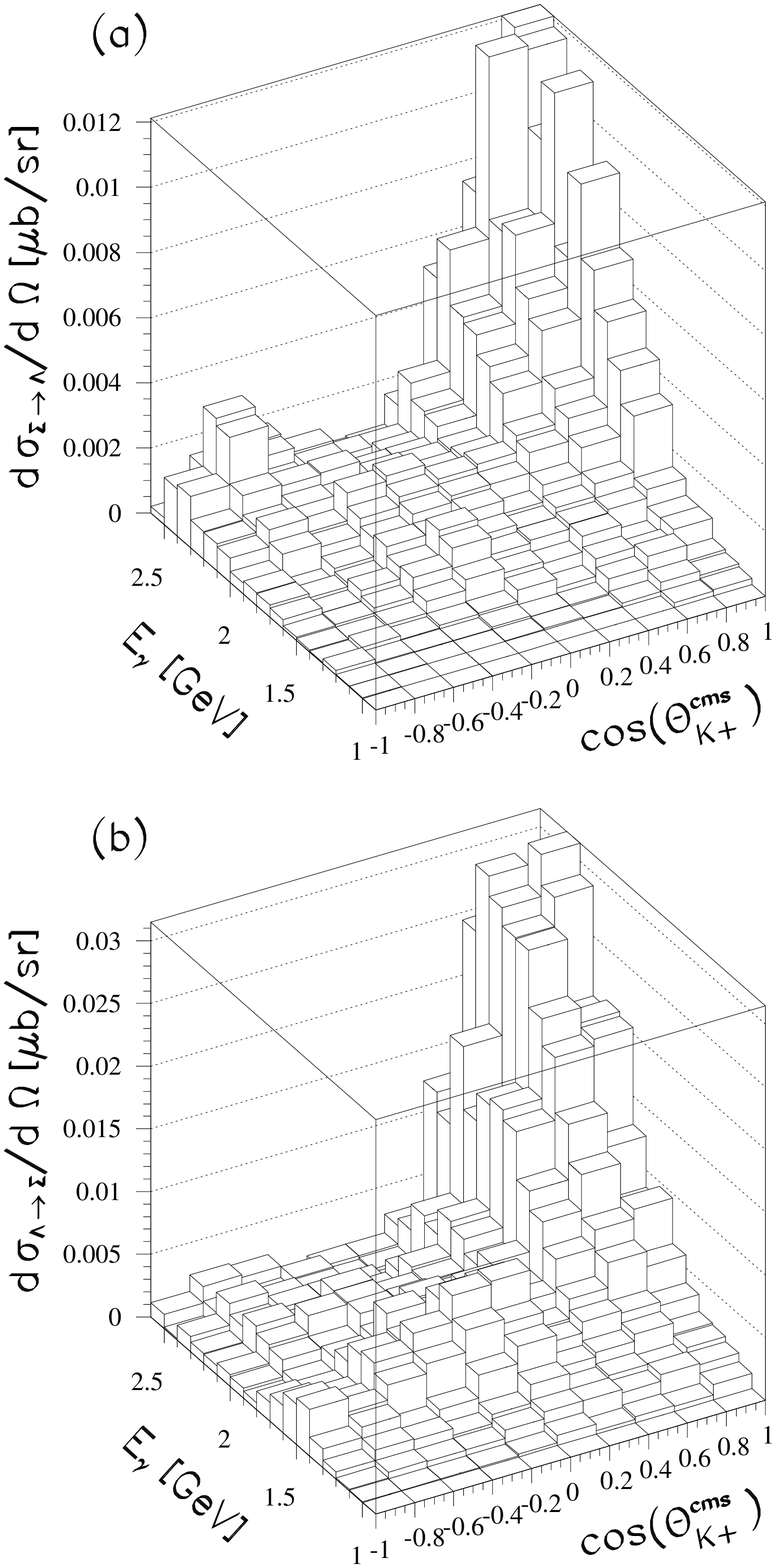}}
\vspace{-0.8cm}
\caption{Differential cross sections of background
due to mutual migrations of \Lambdareaktionvorkomma\ and \Sigmanreaktionvorkomma.
\newline
(a) \Sigmanreaktionvorkomma\ falsely identified as \Lambdareaktionvorkomma.
\newline
(b) \Lambdareaktionvorkomma\ falsely identified as \Sigmanreaktionvorkomma.}
\label{fig:back2}
\vspace{-0.5cm}
\end{figure}

A remaining source of background originates from the migration between
\Lambdareaktionvorkomma\ and \Sigmanreaktionvorkomma\ (see Fig.~\ref{pic:minv2}).
This background was determined from Monte-Carlo generated events and was found
to be on average at the level of 7\% for $K^+\Lambda$ to $K^+\Sigma^0$ and
less than 2\% for $K^+\Sigma^0$ to $K^+\Lambda$ respectively \cite{glander}.
The estimated cross sections corresponding to this background
as a function of photon energy and kaon production angle are shown in Fig.~\ref{fig:back2}.\\
All presented cross sections were corrected for this background in bins of photon energy and
kaon production
\linebreak[4]
angle.

\section{Systematic errors}
\label{sec:systematics}

The error estimates of cross sections shown in figures and tables do not include the overall
systematic uncertainties which are described in the following:

\begin{itemize}
\item The event selection includes a cut on the probability of the kinematic fits which
was set to $P(\chi^2)\,>\,10^{-10}$.
From comparing the probability distribution of experimental and simulated events after the
selection it was estimated that the calculation of the cross section has an overall systematic
normalization uncertainty of about 5\% throughout the energy range~\cite{glander}
by which the true cross sections are systematically underestimated.
\item Background from other reactions was estimated from Monte-Carlo generated events which passed
the same reconstruction and selection criteria as the experimental events.
The relative contribution from reactions which yield non-negligible contributions to the background
are shown in Table~\ref{tab:background}.\\
For the cross section of \Lambdareaktionvorkomma\ the background reactions account for overall
5\% at lower photon energies ($E_\gamma\,<\,1.8~GeV$) and 7\% at higher photon energies ($E_\gamma\,>\,1.8~GeV$).
For \Sigmanreaktionvorkomma\ the corresponding fractions are 7\% and 14\% respectively.
This means that the true cross sections are systematically overestimated by these fractions.\\
\end{itemize}
The systematic shifts of cross sections caused by the two sources concern the overall normalizations.
They are assumed to be independent
from each other and opposite in sign so that they approximately cancel each other except
for \Sigmanreaktionvorkomma\ at the higher photon energies where they combine to a 9\% excess
due to background events.
The errors on cross sections given in figures and tables do not include these normalization errors.

\clearpage

\begin{table*}[htb]
\vspace{-0.2cm}
\caption{Background from other reactions which contribute to the cross sections of the reactions
\Lambdareaktionvorkomma\ and \Sigmanreaktionvorkomma\ in two energy ranges.}
\vspace{0.cm}
\label{tab:background}
\footnotesize
\begin{center}
\begin{tabular}{|c|c|c|c|c|}
\hline
    & \multicolumn{2}{|c|}{\raisebox{-1.01ex}[1.01ex]{\hspace{0.2cm}\Lambdareaktion}} & \multicolumn{2}{|c|}{\raisebox{-1.01ex}[1.01ex]{\Sigmanreaktion}}\\[0.3cm]
%\cline{2-5}
\hline
  Background from & $E_\gamma\,<\,1.8$~GeV & $E_\gamma\,>\,1.8$~GeV & $E_\gamma\,<\,1.8$~GeV & $E_\gamma\,>\,1.8$~GeV\\
\hline
  \Zweipireaktion & $4.5\%\,(\pm\,2.1\%)$ &  $4.7\%\,(\pm\,2.9\%)$ & $2.3\%\,(\pm\,1.8\%)$ &  $1.7\%\,(\pm\,2.4\%)$ \\
  \Dreipireaktion & $0.4\%\,(\pm\,0.6\%)$ &  $2.1\%\,(\pm\,1.5\%)$ & $3.0\%\,(\pm\,1.7\%)$ & $10.6\%\,(\pm\,3.3\%)$ \\
  \Lambdapinull   & $0.0\%\,(\pm\,0.1\%)$ &  $0.0\%\,(\pm\,0.1\%)$ & $1.1\%\,(\pm\,1.0\%)$ &  $1.7\%\,(\pm\,1.3\%)$ \\
  \Lambdaklspip   & $0.0\%\,(\pm\,0.0\%)$ &  $0.0\%\,(\pm\,0.0\%)$ & $0.3\%\,(\pm\,0.6\%)$ &  $0.1\%\,(\pm\,0.3\%)$ \\
\hline
  Total           & $4.9\%\,(\pm\,2.2\%)$ &  $6.8\%\,(\pm\,3.3\%)$ & $6.7\%\,(\pm\,2.8\%)$ & $14.1\%\,(\pm\,4.3\%)$ \\
\hline
\end{tabular}
\end{center}
\vspace{-0.3cm}
\end{table*}

\begin{table*}[htb]
\vspace{-0.2cm}
\caption{Global systematic errors for differential and total cross sections of the reactions
\Lambdareaktionvorkomma\ and \Sigmanreaktionvorkomma\ in two energy ranges. A positive
(negative) background value means that the calculated cross section is underestimated (overestimated)
by the given fraction.}
\vspace{0.cm}
\label{tab:systematics}
\footnotesize
\begin{center}
\begin{tabular}{|c|c|c|c|c|}
\hline
    & \multicolumn{2}{|c|}{\raisebox{-1.01ex}[1.01ex]{\hspace{0.2cm}\Lambdareaktion}} & \multicolumn{2}{|c|}{\raisebox{-1.01ex}[1.01ex]{\Sigmanreaktion}}\\[0.3cm]
\cline{2-5}
                  & $E_\gamma\,<\,1.8$~GeV & $E_\gamma\,>\,1.8$~GeV & $E_\gamma\,<\,1.8$~GeV & $E_\gamma\,>\,1.8$~GeV  \\
\hline
  Probability cut $P_{kin}(\chi^2)\,>\,10^{-10}$   & $+\,5\%$   & $+\,5\%$    & $+\,5\%$   & $+\,5\%$  \\
\hline
  Background from \Zweipireaktion, \Dreipireaktion & $-\,5\%$   & $-\,7\%$    & $-\,7\%$   & $-\,14\%$ \\
  \Lambdapinullvorkomma and \Lambdaklspip          &            &             &            &           \\
\hline
\end{tabular}
\end{center}
\vspace{0.6cm}
\end{table*}

\begin{figure*}[htb]
\vspace{-0.8cm}
\centerline{
\includegraphics[clip,width=1.10\textwidth]{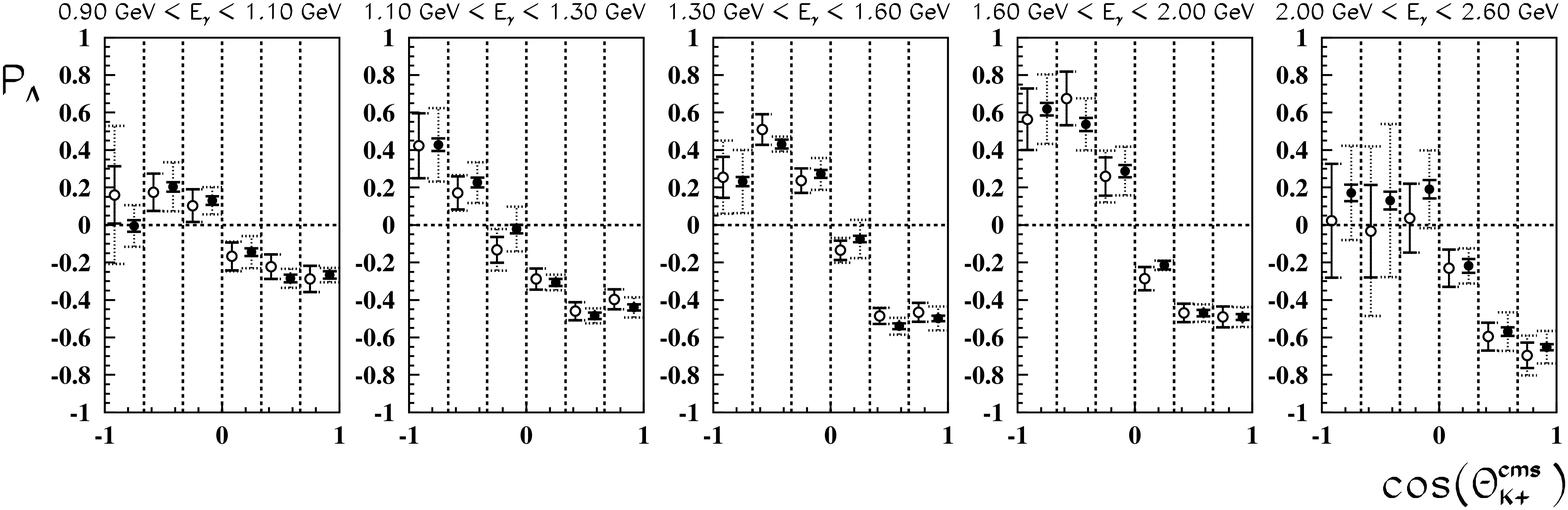}}
\vspace{-0.1cm}
\caption{$\Lambda$ polarizations for \Lambdareaktionvorkomma\ in six bins of the kaon production angle
(separated by the vertical dotted lines) and for five energy bins.
The open circles are the results of fits to the angular distributions as given in the text.
The full circles are results of the up-down asymmetry measurement. The two data sets are
horizontally displaced for visuality.}
\label{fig:pollamsoft}
\vspace{-0.5cm}
\end{figure*}
\begin{figure*}[htb]
\vspace{0.3cm}
\centerline{
\includegraphics[clip,width=1.10\textwidth]{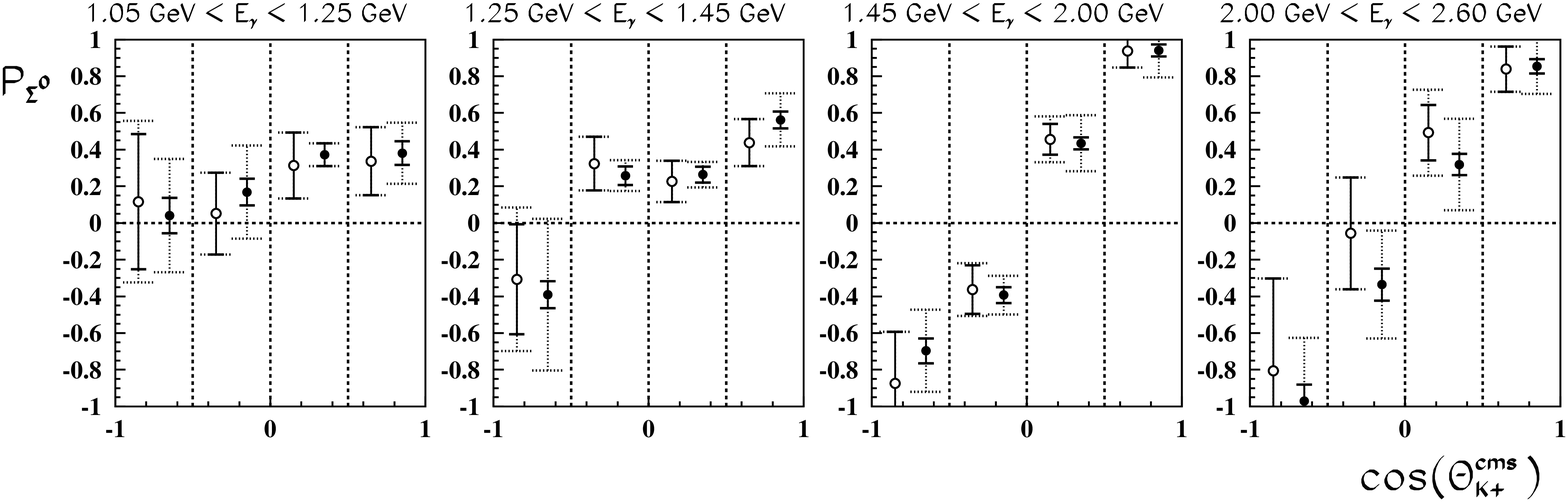}}
\vspace{-0.1cm}
\caption{$\Sigma^0$ polarizations for \Sigmanreaktionvorkomma\ in four bins of the kaon production angle
(separated by the vertical dotted lines) and for four energy bins.
The open circles are the results of fits to the angular distributions as given in the text.
The full circles are results of the up-down asymmetry measurement. The two data sets are
horizontally displaced for visuality.}
\label{fig:polsigsoft}
\vspace{-0.8cm}
\end{figure*}

\clearpage

\section{Hyperon polarizations}

For reactions of the type $a\,+\,b\,\rightarrow\,c\,+\,d$ with unpolarized
initial state particles
and conserved parity the polarization vector, $\vec{P}_d$, in the rest frame of particle $d$
is perpendicular to the production plane \cite{jacob,gottfried}, i.e.\ $\vec{P}\,=\,P\,\vec{n}$,
with $\vec{n}$ being normal to the production plane.
Thus, for \Lambdareaktionvorkomma\ and \Sigmanreaktion
the hyperons $\Lambda$ and $\Sigma^0$ are transversely polarized,
and $P_\Lambda$ and $P_{\Sigma^0}$ denote the polarization parameters.\\
The polarization of $\,\Lambda\,$ in \Lambdareaktionvorkomma\ can be measured
by its parity violating weak decay \Lambdazerfallvorkomma.
The decay angular distribution reads \cite{leeyang}:
\begin{displaymath}
  W(cos\,\theta)\,d\Omega\,=\,\frac{1}{2}\,\Bigl(\,1\,+\,\alpha\,P_\Lambda^{}\,cos\,\theta^{\,}\Bigr)\,d\Omega\,\,\,\,\,\,\,\,\,\,\,\,\,\,\,\,\,\,\,\,,
\end{displaymath}
where $\alpha\,=\,0.642\,\pm\,0.013$ \cite{pdg} is the $\Lambda$ decay parameter
and $\theta$ is defined as the angle between the $\Lambda$ decay proton and $\vec{n}$
in the $\Lambda$~rest frame.\\
For the reaction \Sigmanreaktionvorkomma\ the polarization vector of the $\Sigma^0$ is
related to the polarization vector of the $\Lambda$ produced in the
decay \Sigmazerfall \cite{gatto}:
\begin{displaymath}
  \vec{P}_{\Lambda}\,=\,-\,\Bigl(\vec{P}_{\Sigma^0}\cdot\vec{u}_{\Lambda}\Bigr)\,\vec{u}_{\Lambda}
\end{displaymath}
where $\vec{u}_\Lambda$ is a unit vector describing the direction of
flight of the $\Lambda$ in the $\Sigma^0$ rest frame. Averaging
the event sample over the flight directions of $\Lambda$ yields
\begin{displaymath}
  <\vec{P}_{\Lambda}>_{\vec{u}_{\Lambda}}\,\,=\,\,-\,\frac{1}{3}\,\vec{P}_{\Sigma^0}\,\,\,\,\,\,\,\,\,\,,
\end{displaymath}
i.e. it retains only 1/3 of the sensitivity to the $\Sigma^0$ polarization:
\begin{displaymath}
  W(cos\,\theta)\,\,d\Omega\,=\,\frac{1}{2}\,\left(\,1\,-\frac{1}{3}\,\,\alpha\,P_{\Sigma^0}\,cos\,\theta\right)\,d\Omega
\end{displaymath}
where again $\theta$ is the proton angle in the $\Lambda$ rest frame.\\
The polarization parameters $P_\Lambda$ and $P_{\Sigma^0}$ have been
determined by fits to the angular distributions of the decay proton in bins of kaon production
angle and photon energy.
In a second method $P_\Lambda$ and $P_{\Sigma^0}$ have been determined directly
from the asymmetry of the decay angular distributions according to
\begin{displaymath}
  P_\Lambda\,\,\,=\,+\,\frac{2}{\alpha}\,\frac{N_1\,-\,N_2}
{N_1\,+\,N_2}\,\,\,\,\,\,\,\,\,\,,\,\,\,\,\,\,\,\,\,\,  P_{\Sigma^0}\,=\,-\,\frac{6}{\alpha}\,\frac{N_1\,-\,N_2}
{N_1\,+\,N_2}\,\,\,\,\,\,\,\,\,\,,
\end{displaymath}
with $N_1$ and $N_2$ being the number of events with $cos\,\theta\,>\,0$
and $cos\,\theta\,<\,0$ respectively.\\
Both results are shown in Figures~\ref{fig:pollamsoft}~and~\ref{fig:polsigsoft} and the values
from the fits are given in Appendix~\ref{app:polar}. The inner and outer error bars refer
to $\sigma_w$ and $\sigma_d$ and were determined according to section~\ref{sec:cross}.

\section{Discussion of the results}

\begin{figure}[t]
%\vspace{0.2cm}
\centerline{
\includegraphics[clip,width=0.52\textwidth]{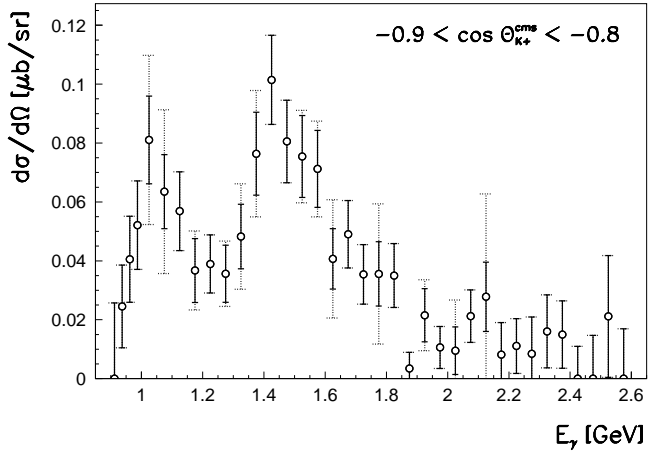}}
\vspace{-0.2cm}
\caption{The differential cross section of \Lambdareaktionvorkomma\ shows an enhancement
for backward produced kaons for photon energies around 1.45~GeV corresponding
to a {\it cms} energy of 1900~MeV.}
\label{fig:wqdiffenhancement}
%\vspace{0.3cm}
\end{figure}

The total cross section of \Lambdareaktionvorkomma\ (see Fig.~\ref{fig:wqtotlambda}) rises
steeply from threshold up to a maximum at a photon energy of about 1.1~GeV. Then, after passing
a nearby flat region around 1.45~GeV, it falls. The gross features of this shape are reflected
in the fitted Legendre coefficients
(see Fig.~\ref{fig:koeffklambda}): The steep rise is connected with $a_0$ together with smaller
contributions of $a_1$ and $a_2$. The coefficient $a_1$ reaches a local minimum around 1.15~GeV,
from where it rises to a local maximum at 1.45~GeV. Towards higher energies, $a_0$ falls
continuously in magnitude while $a_1$ and $a_2$ vary slowly approaching non-zero values.\\
Calculations with effective Lagrangians \cite{coupled1,coupled2,mixedcoupled,janssen} well describe
previous SAPHIR data for the reaction
\Lambdareaktionvorkomma\ by assuming that the production of s- and p-wave resonances
contributes significantly to the cross section close to threshold,
while the exchange of $K^*(892)$ vector and $K_1(1270)$ pseudovector mesons in the t-channel is more
important at higher energies.
All model calculations relate the large s-wave contribution expressed by $a_0$ to the exist\-ence
of the $S_{11}(1650)$ excitation and the p-wave contribution expressed by $a_1$ to
$P_{11}(1710)$ and $P_{13}(1720)$.\\
The structure of the total cross section of \Lambdareaktionvorkomma\ around 1.45~GeV was already observed
in previous
\linebreak[4]
SAPHIR data\ \cite{tran}.
Model calculations which were optimized on these data so far do not give an unambigious picture:
Including these data in their isobar model Bennhold, Mart {\it et al.} interpreted the broad shoulder in
the total cross section at 1.45~GeV as a signal of a $D_{13}(1895)$ resonance
\cite{mixedcoupled,missing1,missing2,missing3}.
Such a $D_{13}$ resonance has been predicted within quark model calculations \cite{capstick,loering}
in this mass range. Janssen {\it et al.} also found that the data description is improved
by including the $D_{13}(1895)$ resonance in their effective Lagrangian approach,
although they point out that also a $P_{13}$, $P_{11}$ and $S_{11}$ in the
same mass region were able to describe the data \cite{janssen}.
Penner and Mosel who carried out a coupled channel analysis \cite{coupled2} obtain a fairly good description of
the data without a significant contribution of
a $D_{13}$ resonance. They explain the structure in the total cross section by a non-resonant
p-wave contribution, stemming from the interference between the s-channel proton and
the t-channel $K^*$ exchange. Saghai also stressed that previous \Lambdareaktionvorkomma\
data can be understood without introducing new resonances \cite{saghai}.\\
In the calculations of Bennhold {\it et al.} an enhancement in the differential cross section
for backward produced kaons in reaction \Lambdareaktion at 1.45~GeV due to the signal of $D_{13}(1895)$
has been predicted~\cite{missing2}. Such an enhancement is
confirmed in the new data around the expected photon energy of 1.45~GeV
as can be seen most significantly for kaon production angles in the range
$-\,0.9\,<\,$\coskaoncmsvorkomma$\,<\,-\,0.8$ (Fig.~\ref{fig:wqdiffenhancement}).
A similar enhancement albeit less pronounced is also observed over the full backward hemisphere.\\
For \Sigmanreaktionvorkomma\ the total cross section (see Fig.~\ref{fig:wqtotsigma}) peaks
around a photon energy of 1.45~GeV. The peak is also seen in the coefficients
$a_0$ and $a_2$ (see Fig.~\ref{fig:koeffksigma0}). The coefficients
$a_1$, $a_3$ and $a_4$ change slope at a photon energy of about 1.8~GeV.
With increasing energy, the contribution of $a_0$ becomes less important while
$a_1$ and, on a smaller scale, also $a_2$ stay with non-zero contributions.\\
Isobar model calculations for \Sigmanreaktionvorkomma\ based on previous
SAPHIR data \cite{mixedcoupled,janssen2}
state that the bump at 1.45~GeV could stem from the $\Delta$ resonances $S_{31}(1900)$ and $P_{31}(1910)$.\\
Above 2~GeV the differential cross sections of both, \Lambdareaktionvorkomma\ and \Sigmanreaktionvorkomma, show
a forward peak with small contributions in the backward hemisphere only.
Apparantly, in this energy range the importance of s-channel resonances decreases and t-channel processes
gain significance.
Kaon photoproduction data for photon energies above 4~GeV are fairly well described by a model
of Guidal {\it et al.}\ \cite{guidal} with reggeized meson propagators. A theoretical understanding of
the transition region itself
is still missing and the data given here should be helpful for this issue.\\
\newline
Both, $\Lambda$ and $\Sigma^0$, are polarized throughout the investigated photon energy range.
The following observations are made (see Figures~\ref{fig:pollamsoft} and~\ref{fig:polsigsoft}):
\begin{itemize}
\item The polarization parameters varies strongly with the production angle of the kaon.
\item They are in general opposite in sign for $\Lambda$ and $\Sigma^0$.
\item The shapes of the angular dependence indicate the same signature throughout the whole energy
range: it tends to be positive for $\Lambda$ (negative for $\Sigma^0$) at backward angles, passes
zero in the central region of
\linebreak[4]
\coskaoncmsohne$\approx 0$ and shows negative (positive) values in
the forward hemisphere.
\item The magnitude of the polarizations parameters varies with energy, for
\Lambdareaktionvorkomma\ less than for \Sigmanreaktionvorkomma.
The $\Sigma^0$ is maximally polarized ($P_{\Sigma^0}\,=\,\pm\,1$) at
forward and backward kaon production angles for photon energies above 1.45~GeV.
\end{itemize}
The opposite signs of the angular dependence of the polarization parameters for $\Lambda$ and $\Sigma^0$
are predicted from SU(6) \cite{thirring} if the same production mechanism
for the s-quark is assumed in both reactions.\\
The persistence of the shape is surprising,
especially for the $\Lambda$ and $\Sigma^0$ production below a photon energy of 1.3~GeV, where
resonance
contributions of $S_{11}(1650)$, $P_{11}(1710)$ and $P_{13}(1720)$ are found to contribute strongly.
Note that the energy bin width for the polarization measurement was chosen to be comparable to the
width of the resonances. 
Current model calculations, which are successful to describe the cross sections
in the ranges of energy and production angle for both reactions of previous data, do not describe the
polarizations of $\Lambda$ and $\Sigma^0$ simultaneously.\\
The polarizations measured for $\Lambda$ and $\Sigma^0$ are similar in shape and magnitude to many
other results obtained with other beams and up to highest energies \cite{paul,heller}, thus suggesting
a general s-quark production scheme \cite{degrand}.

\section{Summary}

New measurements of the reactions \Lambdareaktionvorkomma\ and \Sigmanreaktionvorkomma\
are reported. They were carried out with the SAPHIR detector at ELSA in the photon energy
range from the reaction thresholds to 2.6~GeV.
The results comprise cross sections and hyperon polarizations as a function of the kaon
production angle and the photon energy.
With respect to previous SAPHIR measurements \cite{tran}, the photon energy range was
extended, the differential resolution in both kaon production
angle and photon energy were improved by a factor of two and errors were reduced.\\
The cross sections indicate resonance contributions, for $\Lambda$
production near threshold and for $\Lambda$ and $\Sigma^0$ production
around a photon energy of 1.45~GeV. The hyperons are strongly polarized. The
polarizations for $\Lambda$ and $\Sigma^0$ are in general opposite in sign
along with a pronounced forward-backward asymmetry
which varies slowly with energy.\\

\section{Acknowledgements}

We would like to thank the technical staff of the ELSA machine group for their
invaluable contributions to the
\linebreak[4]
experiment. We gratefully acknowledge the support
by the Deutsche Forschungsgemeinschaft in the framework of the
Schwerpunktprogramm ``Investigations of the hadronic
\linebreak[4]
structure of nucleons and
nuclei with electromagnetic
\linebreak[4]
probes'' (SPP 1034 KL 980/2-3).
\\
\\

\appendix

\section{Differential cross sections of reactions \Lambdareaktionvorkomma\ and \Sigmanreaktion}
\label{app:diff}

The differential cross sections as function of kaon production angle and photon energy
are given as $m\,\pm\,\sigma_d\,(\sigma_w)$.
For definitions of $m$, $\sigma_d$ und $\sigma_w$ see Section~\ref{sec:cross}.
$m\,\pm\,\sigma_d$ should be used for theoretical calculations as $\sigma_d$
accounts, in addition to statistical errors, for the spread of results obtained for the four
data taking periods.
It should be noted that the calculations of cross sections and errors shown in the following
tables do not take into account systematic uncertainties in the overall normalization of cross sections
which were estimated in section~\ref{sec:systematics}.

\clearpage

Sorry, there was not enough space for including the tables for the differential cross sections in this
preprint. Ask for theses by email to glander@physik.uni-bonn.de or see the paper expected to be published soon.

\begin{table}[htb]
%\vspace{-0.2cm}
\caption{Differential cross section $\frac{d\,\sigma}{d\,\Omega}~[\mbox{$\mu b/sr$}]$ of reaction \Lambdareaktion
for photon energies $0.9~\mbox{GeV}\,\le\,E_\gamma\,\le\,1.4~\mbox{GeV}$.}
\label{tab:appwqdifflambda1}
\end{table}
\begin{table}[htb]
\vspace{-0.2cm}
\caption{Differential cross section $\frac{d\,\sigma}{d\,\Omega}~[\mbox{$\mu b/sr$}]$ of reaction \Lambdareaktion
for photon energies $1.4~\mbox{GeV}\,\le\,E_\gamma\,\le\,2.0~\mbox{GeV}$.}
\label{tab:appwqdifflambda2}
\end{table}
\begin{table}[htb]
\vspace{-0.2cm}
\caption{Differential cross section $\frac{d\,\sigma}{d\,\Omega}~[\mbox{$\mu b/sr$}]$ of reaction \Lambdareaktion
for photon energies $2.0~\mbox{GeV}\,\le\,E_\gamma\,\le\,2.6~\mbox{GeV}$.}
\label{tab:appwqdifflambda3}
\end{table}

\begin{table}[htb]
\vspace{-0.2cm}
\caption{Differential cross section $\frac{d\,\sigma}{d\,\Omega}~[\mbox{$\mu b/sr$}]$ of reaction \Sigmanreaktion
for photon energies $1.05~\mbox{GeV}\,\le\,E_\gamma\,\le\,1.55~\mbox{GeV}$.}
\label{tab:appwqdiffsigma1}
\end{table}
\begin{table}[htb]
\vspace{-0.2cm}
\caption{Differential cross section $\frac{d\,\sigma}{d\,\Omega}~[\mbox{$\mu b/sr$}]$ of reaction \Sigmanreaktion
for photon energies $1.55~\mbox{GeV}\,\le\,E_\gamma\,\le\,2.15~\mbox{GeV}$.}
\label{tab:appwqdiffsigma2}
\end{table}
\begin{table}[htb]
\vspace{-0.2cm}
\caption{Differential cross section $\frac{d\,\sigma}{d\,\Omega}~[\mbox{$\mu b/sr$}]$ of reaction \Sigmanreaktion
for photon energies $2.15~\mbox{GeV}\,\le\,E_\gamma\,\le\,2.60~\mbox{GeV}$.}
\label{tab:appwqdiffsigma3}
\end{table}

\clearpage

\section{Hyperon polarizations}
\label{app:polar}

The polarizations are given as $m\,\pm\,\sigma_d\,(\sigma_w)$.
For definitions of $m$, $\sigma_d$ und $\sigma_w$ see Section~\ref{sec:cross}.
$m\,\pm\,\sigma_d$ should be used for theoretical calculations.

\begin{table}[htb]
\vspace{0.2cm}
\caption{\small Hyperon polarization $P_\Lambda$ for reaction \Lambdareaktion
for photon energies $0.9~GeV\,\le\,E_\gamma\,\le\,2.6~GeV$. \normalsize}
\vspace{0.cm}
\label{tab:apppolnewbinslambdagerade1}
\scriptsize
\begin{center}
\begin{tabular}{|rcr|r|r|r|r|}
\hline
%\multicolumn{7}{|c|}{\raisebox{-1.01ex}[1.01ex]{\small Hyperonpolarisation $P_\Lambda$ in der Reaktion \Lambdareaktion}} \\[0.3cm]
%\hline
\multicolumn{3}{|r|}{\raisebox{-1.01ex}[1.01ex]{\normalsize \hspace{0.2cm}\coskaoncmsohne}} & \multicolumn{4}{|c|}{\raisebox{-1.01ex}[1.01ex]{\normalsize $E_\gamma~[GeV]$}} \scriptsize \\[0.3cm]
\cline{4-7}
\multicolumn{3}{|r|}{ } & \multicolumn{1}{|c|}{0.90\,-\,1.10} & \multicolumn{1}{|c|}{1.10\,-\,1.30} & \multicolumn{1}{|c|}{1.30\,-\,1.60} & \multicolumn{1}{|c|}{1.60\,-\,2.00}\\
\hline
$-1$   & $\ldots$ & $-2/3$ & $0.161\,\pm\,0.369\,(0.153)$ & $0.423\,\pm\,0.173\,(0.173)$ & $0.255\,\pm\,0.195\,(0.109)$ & $0.564\,\pm\,0.165\,(0.165)$ \\
$-2/3$ & $\ldots$ & $-1/3$ & $0.175\,\pm\,0.099\,(0.099)$ & $0.171\,\pm\,0.094\,(0.088)$ & $0.510\,\pm\,0.082\,(0.082)$ & $0.675\,\pm\,0.143\,(0.143)$ \\
$-1/3$ & $\ldots$ & $0$    & $0.103\,\pm\,0.087\,(0.087)$ & $-0.133\,\pm\,0.112\,(0.069)$ & $0.237\,\pm\,0.065\,(0.065)$ & $0.259\,\pm\,0.138\,(0.103)$ \\
$0$    & $\ldots$ & $1/3$  & $-0.168\,\pm\,0.079\,(0.075)$ & $-0.288\,\pm\,0.056\,(0.056)$ & $-0.135\,\pm\,0.067\,(0.051)$ & $-0.287\,\pm\,0.062\,(0.062)$ \\
$1/3$  & $\ldots$ & $2/3$  & $-0.223\,\pm\,0.066\,(0.066)$ & $-0.460\,\pm\,0.048\,(0.048)$ & $-0.485\,\pm\,0.043\,(0.043)$ & $-0.469\,\pm\,0.049\,(0.049)$ \\
$2/3$  & $\ldots$ & $1$    & $-0.288\,\pm\,0.070\,(0.070)$ & $-0.397\,\pm\,0.053\,(0.053)$ & $-0.466\,\pm\,0.051\,(0.051)$ & $-0.491\,\pm\,0.056\,(0.056)$ \\
\hline
\multicolumn{3}{|r|}{\raisebox{-1.01ex}[1.01ex]{\normalsize \hspace{0.2cm}\coskaoncmsohne}} & \multicolumn{4}{|c|}{\raisebox{-1.01ex}[1.01ex]{\normalsize $E_\gamma~[GeV]$}} \scriptsize \\[0.3cm]
\cline{4-7}
\multicolumn{3}{|r|}{ } & \multicolumn{1}{|c|}{2.00\,-\,2.60} \\
\cline{1-4}
$-1$   & $\ldots$ & $-2/3$ & $0.023\,\pm\,0.304\,(0.304)$ \\
$-2/3$ & $\ldots$ & $-1/3$ & $-0.033\,\pm\,0.452\,(0.247)$ \\
$-1/3$ & $\ldots$ & $0$    & $0.036\,\pm\,0.184\,(0.184)$ \\
$0$    & $\ldots$ & $1/3$  & $-0.231\,\pm\,0.100\,(0.100)$ \\
$1/3$  & $\ldots$ & $2/3$  & $-0.596\,\pm\,0.074\,(0.074)$ \\
$2/3$  & $\ldots$ & $1$    & $-0.696\,\pm\,0.106\,(0.068)$ \\
\cline{1-4}
\end{tabular}
\end{center}
\vspace{-0.4cm}
\end{table}

\begin{table}[htb]
\vspace{0.2cm}
\caption{\small Hyperon polarization $P_{\Sigma^0}$ in reaction \Sigmanreaktion
for photon energies $1.05~GeV\,\le\,E_\gamma\,\le\,2.60~GeV$. \normalsize}
\vspace{0.cm}
\label{tab:apppololdbinssigmagerade1}
\scriptsize
\begin{center}
\begin{tabular}{|rcr|r|r|r|r|}
%\hline
%\multicolumn{7}{|c|}{\raisebox{-1.01ex}[1.01ex]{\small Hyperonpolarisation $P_{\Sigma^0}$ in der Reaktion \Sigmanreaktion}} \\[0.3cm]
\hline
\multicolumn{3}{|r|}{\raisebox{-1.01ex}[1.01ex]{\normalsize \hspace{0.2cm}\coskaoncmsohne}} & \multicolumn{4}{|c|}{\raisebox{-1.01ex}[1.01ex]{\normalsize $E_\gamma~[GeV]$}} \scriptsize \\[0.3cm]
\cline{4-7}
\multicolumn{3}{|r|}{ } & \multicolumn{1}{|c|}{1.05\,-\,1.25} & \multicolumn{1}{|c|}{1.25\,-\,1.45} & \multicolumn{1}{|c|}{1.45\,-\,2.00} & \multicolumn{1}{|c|}{2.00\,-\,2.60}\\
\hline
$-1$   & $\ldots$ & $-1/2$ & $~\,~0.116\,\pm\,0.440\,(0.369)$ & $-0.307\,\pm\,0.391\,(0.300)$ & $-0.874\,\pm\,0.280\,(0.280)$ & $-0.806\,\pm\,0.504\,(0.504)$ \\
$-1/2$ & $\ldots$ & $0$    & $~\,~0.052\,\pm\,0.223\,(0.223)$ & $0.324\,\pm\,0.146\,(0.146)$ & $-0.363\,\pm\,0.143\,(0.132)$ & $-0.057\,\pm\,0.304\,(0.304)$ \\
$0$    & $\ldots$ & $1/2$  & $~\,~0.313\,\pm\,0.180\,(0.180)$ & $0.226\,\pm\,0.113\,(0.113)$ & $0.456\,\pm\,0.125\,(0.084)$ & $0.493\,\pm\,0.234\,(0.151)$ \\
$1/2$  & $\ldots$ & $1$    & $~\,~0.336\,\pm\,0.185\,(0.185)$ & $0.438\,\pm\,0.128\,(0.128)$ & $0.937\,\pm\,0.089\,(0.089)$ & $0.839\,\pm\,0.123\,(0.123)$ \\
\hline
\end{tabular}
\end{center}
\vspace{-0.4cm}
\end{table}

\clearpage

%\section*{References}

%%% Local Variables: ***
%%% mode: LaTeX ***
%%% ispell-dictionary:"deutsch-latin1"***
%%% mode: iso-accents ***
%%% TeX-master: "dok.tex"***
%%% End: ***


\begin{thebibliography}{99}
%
\bibitem{glander}
K.-H.\,Glander: doctoral thesis Bonn 2003, BONN-IR-2003-05; to be found on the web:\\
http://saphir.physik.uni-bonn.de/saphir/thesis.html
%hss.ulb.uni-bonn.de/ulb_bonn/diss_online/math_nat_fak/2003/glander_karl_heinz
%
\bibitem{capstick}
S.\,Capstick, W.\,Roberts: Phys. Rev. D {\bf 58}, 074011 (1998)
%
\bibitem{loering}
U.\,L{\"o}ring, B.\,Ch.\,Metsch, H.\,R.\,Petry: Eur. Phys. J. A {\bf 10}, 395 (2001)
%
\bibitem{schwille}
W.\,J.\,Schwille {\it et al.}: Nucl. Instr. Meth. A {\bf 344}, 470 (1994)
%
\bibitem{husmann}
D.\,Husmann, W.\,J.\,Schwille: Phys. Bl. {\bf 44}, 40 (1988)
%
\bibitem{tran}
M.\,Q.\,Tran {\it et al.}: Phys. Lett. B {\bf 445}, 20 (1998)
%
\bibitem{topas2}
%J.\,Link: Diplomarbeit Bonn 1994, Bonn IR-94-33; doktoral thesis Bonn 2000, ISKP-Report 1/2000
R.\,Burgwinkel: doctoral thesis Bonn 1996, Bonn IR-96-02
%
\bibitem{barth2}
J.\,Barth: doctoral thesis Bonn 2002, BONN-IR-02-6
%
\bibitem{pdg}
Particle Data Group (K. Hagiwara {\it et al.}): Phys. Rev. D {\bf 66}, 010001 (2002)
%
\bibitem{geant}
R.~Brun, F.~Carena {\it et al.}: GEANT~Simulating Program for Particle Physics Experiments, Version
2.0, CERN~DD/US/86
%
\bibitem{omega}
J.\,Barth {\it et al.}: Low-energy photoproduction of $\omega$-mesons,
accepted for publication in Eur. Phys. J.
%
\bibitem{hannappeld}
J.\,Hannappel: doctoral thesis Bonn 1997, BONN-IR-97-15
%
\bibitem{neuerburgd}
W.\,Neuerburg: doctoral thesis Bonn 1999, BONN-IR-99-06
%
%\bibitem{rolandiblum}
%W.\,Blum, L.\,Rolandi: Particle Detection with Drift Chambers, Springer-Verlag, 1993
%
\bibitem{landolt}
Landolt-B\"ornstein:
Numerical Data and Functional Relationships in Science and Technology,
New Series I/12b (1988)
%
\bibitem{phi}
J.\,Barth {\it et al.}: Low-energy photoproduction of $\Phi$-mesons, Eur. Phys. J. A {\bf 17}, 269-274 (2003)
%
\bibitem{rho}
C.\,Wu {\it et al.}: $\rho$ photoproduction at SAPHIR, in preparation
%
\bibitem{ABBHHM}
ABBHHM collaboration: Phys. Rev. {\bf 188}, 2060 (1969)
%
\bibitem{jacob}
M.\,Jacob, G.\,C.\,Wick: Annals of Physics {\bf 7}, 404 (1959)
%``On the General Theory of Collisions for Particles with Spin''\\
%
\bibitem{gottfried}
K.\,Gottfried, J.\,D.\,Jackson: Nuovo Cimento {\bf 33}, 309 (1964)
%``On the Connection between Production Mechanism and Decay of Resonances at High Energies''\\
%
\bibitem{leeyang}
T.\,D.\,Lee, C.\,N.\,Yang: Phys. Rev. {\bf 108}, 1645 (1957)
%``General Partial Wave Analysis of the Decay of Hyperon of Spin 1/2''\\
%
\bibitem{gatto}
R.\,Gatto: Phys. Rev. {\bf 109}, 610 (1958)
%``Relations between the Hyperon Polarizations in Associated Production''\\
%
%\bibitem{david_saghai}
%J.\,C.\,David {\it et al.}: Phys. Rev. C {\bf 53}, 2613 (1996)
%
%\bibitem{steininger}
%S.\,Steininger {\it et al.},
%{\it Physics Letters } {\bf B 391}(1997) 446
%
%\bibitem{kaiser}
%N. Kaiser {\it et al.},
%{\it Nucl. Phys. } {\bf A 612} (1997) 297
%
%\bibitem{feuster}
%T.\,Feuster and U.\,Mosel: Phys. Rev. C {\bf 58}, 457 (1998);
%
\bibitem{coupled1}
T.\,Feuster and U.\,Mosel: Phys. Rev. C {\bf 59}, 460 (1999)
%; {\it Preprint} nucl-th/9803057 (1998)
%''Photon- and meson-induced reactions on the nucleon''
%
\bibitem{coupled2}
G.\,Penner and U.\,Mosel: Phys. Rev. C {\bf 66}, 055212 (2002)
%; {\it Preprint} nucl-th/0207069 (2002)
%''Vector meson production and nucleon resonance analysis in a coupled-channel approach for energies $m_N\,<\,\sqrt{s}\,<\,2~GeV$; II: photon-induced reactions''
%
\bibitem{mixedcoupled}
C.\,Bennhold, T.\,Mart, A.\,Waluyo, H.\,Haberzettl, G.\,Penner, T.\,Feuster, U.\,Mosel: {\it Preprint} nucl-th/9901066
%''Nucleon Resonances in Kaon Photoproduction''
%
\bibitem{janssen}
S.\,Janssen, J.\,Ryckebusch, W.\,Van\,Nespen, D.\,Debruyne, T.\,Van\,Cauteren: Eur. Phys. J. A {\bf 11}, 105 (2001)
%; {\it Preprint} nucl-th/0105008
%
\bibitem{missing1}
T.\,Mart and C.\,Bennhold: Phys. Rev. C {\bf 61}, (R)012201 (2000)
%; {\it Preprint} nucl-th/9906096
% Evidence for a missing nucleon resonance in kaon photoproduction
%
\bibitem{missing2}
C.\,Bennhold, H.\,Haberzettl, T.\,Mart: {\it Proceedings of the Second
International Conference on Perspectives in Hadronic Physics}, edited by Sigfrido Boffi, Claudio Ciofi degli Atti \& Mauro Giannini
(World Scientific, 1999)
%; {\it Preprint} nucl-th/9909022
%``A New Resonance in $K^+ \Lambda$ Electroproduction: The $D_{13}(1895)$ And Its
%Electromagnetic Form Factors''\\
%
\bibitem{missing3}
C.\,Bennhold, A.\,Waluyo, H.\,Haberzettl, T.\,Mart, G.\,Penner, U.\,Mosel: {\it Preprint} nucl-th/0008024
%``Missing Nucleon Resonances In Kaon Production With Pions And Photons''
%
\bibitem{saghai}
B.\,Saghai: {\it Preprint} nucl-th/0105001
%
\bibitem{janssen2}
S.\,Janssen, J.\,Ryckebusch, D.\,Debruyne, T.\,Van\,Cauteren: Phys. Rev. C {\bf 66}, 035202 (2002)
%; {\it Preprint} nucl-th/0105008
%\bibitem{mart}
%older model !!! S. z.B. Verweis [9] in Preprint nucl-th/9901066
%T.\,Mart, C.\,Bennhold, and C.\,E.\,Hyde-Wright: 
%Phys. Rev. C {\bf 51}, R1074 (1995);
%T.\,Mart and C.\,Bennhold: Few-Body Syst. Suppl. {\bf 9}, 369 (1995);
%C.\,Bennhold, T.\,Mart, and D.\,Kusno in {\it Proceedings of the 4th CEBAF/INT
%Workshop on $N^*$ Physics}, edited by T.-S.\,H.\,Lee and W.\,Roberts (World
%Scientific, Singapore, 1997), p. 166
%
\bibitem{guidal}
M.\,Guidal, J.-M.\,Laget, M.\,Vanderhaeghen:
Nucl. Phys. A {\bf 627}, 645 (1997);
%M.\,Vanderhagen, M.\,Guidal, J.\,M.\,Laget:
%Phys. Rev. C {\bf 57}, 1454 (1998);
%M.\,Guidal:
%{\it These}
%(DAPNIA/SPhN-96-03T, 1/1997), unpublished
%
\bibitem{thirring}
See e.g.~W.\,Thirring:
% {\it Electromagnetic Properties of Hadrons in the Static $SU_6$ Modell},
Acta Physica Austriaca {\bf Sup. II} (1966)
%
\bibitem{paul}
E.\,Paul: Italian Phys. Society 1992, Vol. 44, Proc. of the Conference on THE ELFE PROJECT,
Mainz 1992, p. 379
%
\bibitem{heller}
K.\,Heller: Proc. of the 9th International Symposion on High Energy Spin Physics, Bonn 1990,
Springer ISBN 3-540-54127-6
%
\bibitem{degrand}
Th.\,A.\,DeGrand and H.\,Miettinen: Phys. Rev. D {\bf 24}, 2419 (1981)
%
\end{thebibliography}
\end{document}